\documentclass[12pt]{article}
\pdfoutput=1

\usepackage{amsmath}
\usepackage{amssymb}
\usepackage{epsfig}
\usepackage{epstopdf}
\usepackage{latexsym}
\usepackage{color}
\numberwithin{equation}{section}
\usepackage[
      colorlinks=false,
      linkcolor=darkblue,  
      urlcolor=blue,    
      filecolor=blue,     
      citecolor=red,
linktocpage=true,
      pdfstartview=FitV,
      bookmarksopen=true    
      ]{hyperref}

\usepackage{amssymb}
\usepackage[bbgreekl]{mathbbol}
\DeclareSymbolFontAlphabet{\mathbbl}{bbold}

\begin{document}

\begin{titlepage}

\centerline
\centerline
\centerline
\bigskip
\bigskip
\centerline{\Huge \rm Spindle black holes and mass-deformed ABJM}
\bigskip
\bigskip
\bigskip
\bigskip
\bigskip
\bigskip
\bigskip
\centerline{\rm Minwoo Suh}
\bigskip
\centerline{\it School of General Education, Kumoh National Institute of Technology,}
\centerline{\it Gumi, 39177, Korea}
\bigskip
\centerline{\tt minwoosuh1@gmail.com} 
\bigskip
\bigskip
\bigskip
\bigskip
\bigskip
\bigskip
\bigskip

\begin{abstract}
\noindent By extending the method developed by Arav, Gauntlett, Roberts and Rosen, we construct supersymmetric $AdS_2\times\mathbbl{\Sigma}$ solutions of gauged $\mathcal{N}=8$ supergravity which are asymptotic to the $SU(3)\times{U}(1)$-invariant Warner fixed point, where $\mathbbl{\Sigma}$ is a spindle. The Warner fixed point is dual to the mass-deformed ABJM theory.  The solutions are in the anti-twist class. We numerically calculate the Bekenstein-Hawking entropy of the presumed black holes with the $AdS_2\times\mathbbl{\Sigma}$ horizon.
\end{abstract}

\vskip 6cm

\flushleft {May, 2024}

\end{titlepage}

\tableofcontents

\section{Introduction}

Topological twist is a systematic way of realizing supersymmetry in field theory, \cite{Witten:1988ze}, and in gravity, \cite{Maldacena:2000mw}, with a wide range of applications. Recently, a novel class of anti-de Sitter solutions were discovered which realize supersymmetry in different ways from the topological twist. The solutions are obtained from branes wrapping an orbifold, namely, a spindle, \cite{Ferrero:2020laf}. The spindle, $\mathbbl{\Sigma}$, is an orbifold, $\mathbbl{WCP}_{[n_-,n_+]}^1$, with conical deficit angles at two poles. The spindle numbers, $n_-$, $n_+$, are arbitrary coprime positive integers. The solutions were first constructed from D3-branes, \cite{Ferrero:2020laf, Hosseini:2021fge, Boido:2021szx}, and then applied to other branes: M2-branes, \cite{Ferrero:2020twa, Cassani:2021dwa, Ferrero:2021ovq, Couzens:2021rlk}, M5-branes, \cite{Ferrero:2021wvk}, D4-branes, \cite{Faedo:2021nub, Giri:2021xta}, mass-deformed D3-branes, \cite{Arav:2022lzo}, and D2-branes, \cite{Couzens:2022yiv}. Furthermore, two possible ways of realizing supersymmetry, topologically topological twist and anti-twist, were classified, \cite{Ferrero:2021etw, Couzens:2021cpk}.

The orbifold solutions with a single conical deficit angle, namely, a half spindle, is topologically a disk. The $AdS_5$ solutions from M5-branes wrapped on a topological disk were first constructed in  \cite{Bah:2021mzw, Bah:2021hei}, and were proposed to be the gravity dual to a class of 4d $\mathcal{N}=2$ Argyres-Douglas theories, \cite{Argyres:1995jj}. See also \cite{Couzens:2022yjl, Bah:2022yjf} for further studies. Solutions from other branes wrapped on a topological disk were constructed: D3-branes, \cite{Couzens:2021tnv, Suh:2021ifj}, M2-branes, \cite{Suh:2021hef, Couzens:2021rlk}, and D4-branes, \cite{Suh:2021aik}. The disk solutions are locally identical to spindle solutions, but originate from a different global completion. See also \cite{Karndumri:2022wpu} for more disk solutions and \cite{Gutperle:2022pgw} for defect solutions from the other different completion of global solutions.

The line of study was soon generalized to branes wrapping higher-dimensional orbifolds, in particular, four-dimensional orbifolds: the direct product of a spindle and a constant curvature surface. M5-branes on $\mathbbl{\Sigma}\times\Sigma_{\mathfrak{g}}$, \cite{Boido:2021szx, Suh:2022olh}, D4-branes on $\mathbbl{\Sigma}\times\Sigma_{\mathfrak{g}}$, \cite{Giri:2021xta, Faedo:2021nub, Suh:2022olh}, and M5-branes on $\mathbbl{\Sigma}\times{H}^3$, \cite{Couzens:2022yiv}, were constructed where $\mathbbl{\Sigma}$ is a spindle or a disk and $\Sigma_{\mathfrak{g}}$ is a Riemann surface of genus, $\mathfrak{g}$. Furthermore, branes wrapped on a spindle fibered over another spindle or a Riemann surface were also constructed. In \cite{Cheung:2022ilc} M5-branes wrapped on $\mathbbl{\Sigma}_1\ltimes\mathbbl{\Sigma}_2$ and $\mathbbl{\Sigma}\ltimes\Sigma_{\mathfrak{g}}$ were constructed. Then, D4-branes wrapped on $\mathbbl{\Sigma}_1\ltimes\mathbbl{\Sigma}_2$ and $\mathbbl{\Sigma}\ltimes\Sigma_{\mathfrak{g}}$ were found in \cite{Couzens:2022lvg, Faedo:2022rqx}. See \cite{Faedo:2021nub, Faedo:2022rqx, Boido:2022iye, Boido:2022mbe} also for the recent development of gravitational blocks, \cite{Hosseini:2019iad}, and entropy functions for spindle solutions.

A common feature of the examples we have listed is that they could asymptote to the maximally supersymmetric vacuum in their respective dimensions. Via the AdS/CFT correspondence, \cite{Maldacena:1997re}, they are dual to 3d ABJM theory, 4d $\mathcal{N}=4$ super Yang-Mills theory, 5d $USp(2N)$ gauge theory and 6d $\mathcal{N}=(2,0)$ theory, respectively. However, in \cite{Arav:2022lzo} by Arav, Gauntlett, Roberts and Rosen, it was shown that we can construct orbifold solutions asymptote to the $AdS$ vacua with less supersymmetries: $AdS_3\times\mathbbl{\Sigma}$ solution asymptotic to the $\mathcal{N}=2$ $AdS_5$ vacuum, \cite{Khavaev:1998fb, Freedman:1999gp, Pilch:2000fu}, dual to the Leigh-Strassler SCFTs, \cite{Leigh:1995ep}, was constructed. Furthermore, holographically, central charge was computed and matched with the field theory result from the Leigh-Strassler theory on $\mathbb{R}^{1,1}\times\mathbbl{\Sigma}$.

In this work, we construct supersymmetric $AdS_2\times\mathbbl{\Sigma}$ solutions asymptotic to the $\mathcal{N}=2$ $AdS_4$ vacuum, \cite{Warner:1983vz, Warner:1983du}, which is dual to the mass-deformed ABJM theory, \cite{Benna:2008zy, Klebanov:2008vq}. The holographic RG flow from the $\mathcal{N}=8$ $AdS_4$ vacuum dual to the ABJM theory to the $\mathcal{N}=2$ $AdS_4$ vacuum, \cite{Warner:1983vz, Warner:1983du}, dual to the mass-deformed ABJM was constructed in \cite{Ahn:2000aq, Ahn:2000mf} and uplifted to eleven-dimensional supergravity in \cite{Corrado:2001nv}. In light of the discovery of ABJM theory, \cite{Aharony:2008ug}, the mass-deformed ABJM theory was further understood in \cite{Benna:2008zy, Klebanov:2008vq, Bobev:2009ms}. See \cite{Bobev:2018wbt} also for the gravity calculation of three-sphere free energy of mass-deformed ABJM theory. The supersymmetric black hole solutions interpolating the $\mathcal{N}=2$ $AdS_4$ vacuum and the horizon of $AdS_2\times\Sigma_{\mathfrak{g}}$ was constructed in \cite{Bobev:2018uxk}. 

We employed the $U(1)^2$-invariant truncation, \cite{Bobev:2018uxk}, of $SO(8)$-gauged $\mathcal{N}=8$ supergravity in four dimensions, \cite{deWit:1982bul}, which is a consistent truncation of eleven-dimensional supergravity, \cite{Cremmer:1978km}, on a seven-dimensional sphere, \cite{deWit:1986oxb}. We derive the BPS equations for the $AdS_2\times\mathbbl{\Sigma}$ solutions asymptotic to the $\mathcal{N}=2$ $AdS_4$ vacuum dual to the mass-deformed ABJM theory. We study the BPS equations and find the flux quantizations through the spindle, $\mathbbl{\Sigma}$. We find all the necessary algebraic constraints to determine the boundary conditions to construct solutions explicitly. Due to the complexity of the expressions of constraints, we cannot solve for the boundary conditions analytically. We resort to the numerical determination of boundary conditions for each choice of the spindle numbers, $n_{N,S}$, $t_{N,S}$, and the flavor charges, $p_{F_1}$, $p_{F_2}$. Using these boundary conditions, we explicitly construct the solutions numerically. We also numerically calculate the Bekenstein-Hawking entropy of the presumed black hole with the spindle horizon. As a check, for the case of no flavor symmetries, $p_{F_1}=p_{F_2}=0$, the numerical values of Bekenstein-Hawking entropy precisely match with the one of solutions from minimal gauged supergravity, \cite{Ferrero:2020twa}.

For the structure of the work, we will closely follow \cite{Arav:2022lzo} as it is well organized and also to facilitate the comparison.

In section \ref{sec2} we review the $U(1)^2$-invariant truncation. In section \ref{sec3} we study the BPS equations and calculate the Bekenstein-Hawking entropy. In section \ref{sec4}, we numerically construct the solutions. In section \ref{sec5} we conclude. In appendix \ref{appA} we review the construction of the $U(1)^2$-invariant truncation of gauged $\mathcal{N}=8$ supergravity. In appendix \ref{appB} we present the derivation of the BPS equations.

\section{The supergravity model} \label{sec2}

We consider the $U(1)^2$-invariant truncation of $SO(8)$-gauged $\mathcal{N}=8$ supergravity in four dimensions, \cite{deWit:1982bul}, which was studied in \cite{Bobev:2018uxk}. See appendix \ref{appA} for details. The field content is the metric, four $U(1)$ gauge fields, $A^\alpha$, $\alpha\,=\,0,\,\ldots,\,3$, and four complex scalar fields, $(\chi,\psi)$ and $(\lambda_i,\varphi_i)$, $i\,=\,1,\,\ldots,\,3$. We set $\varphi_i\,=\,\pi$. We employ mostly plus signature. The bosonic Lagrangian of the truncation is given by
\begin{align} \label{mlag}
&e^{-1}\mathcal{L}\,=\,\frac{1}{2}R-\partial_\mu\chi\partial^\mu\chi-\frac{1}{4}\sinh^2\left(2\chi\right)D_\mu\psi{D}^\mu\psi-\sum_{i=1}^3\partial_\mu\lambda_i\partial^\mu\lambda_i-g^2\mathcal{P} \notag \\
&-\frac{1}{4}\left[e^{-2\left(\lambda_1+\lambda_2+\lambda_3\right)}F_{\mu\nu}^0F^{0\mu\nu}+e^{-2\left(\lambda_1-\lambda_2-\lambda_3\right)}F_{\mu\nu}^1F^{1\mu\nu}+e^{2\left(\lambda_1-\lambda_2+\lambda_3\right)}F_{\mu\nu}^2F^{2\mu\nu}+e^{2\left(\lambda_1+\lambda_2-\lambda_3\right)}F_{\mu\nu}^3F^{3\mu\nu}\right]\,,
\end{align}
where we define
\begin{equation}
D\psi\,\equiv\,d\psi-g\left(A^0-A^1-A^2-A^3\right)\,.
\end{equation}
The scalar potential is given by
\begin{equation}
\mathcal{P}\,=\,\frac{1}{2}\left(\frac{\partial{W}}{\partial\chi}\right)^2+\frac{1}{2}\sum_{i=1}^3\left(\frac{\partial{W}}{\partial\lambda_i}\right)^2-\frac{3}{2}W^2\,,
\end{equation}
where the superpotential is
\begin{equation}
W\,=\,\frac{2e^{2\left(\lambda_1+\lambda_2+\lambda_3\right)}\sinh^2\chi-\cosh^2\chi\left(e^{2\left(\lambda_1+\lambda_2+\lambda_2\right)}+e^{2\lambda_1}+e^{2\lambda_2}+e^{2\lambda_3}\right)}{2e^{\lambda_1+\lambda_2+\lambda_3}}\,.
\end{equation}
The R-symmetry vector field, the massive vector field, and two flavor symmetry vector fields are given by, respectively,
\begin{align} \label{fourvectors}
A_R\,=&\,-g\left(A^0+A^1+A^2+A^3\right)\,, \notag \\
A_m\,=&\,g\left(A^0-A^1-A^2-A^3\right)\,, \notag \\
A_{F_1}\,=&\,\frac{1}{2}g\left(A^1-A^2\right)\,, \notag \\
A_{F_2}\,=&\,\frac{1}{\sqrt{3}}g\left(A^1+A^2-2A^3\right)\,.
\end{align}

The BPS equations are obtained by setting the fermionic supersymmetry variations to vanish. The spin-3/2 field variations reduce to
\begin{equation}
\left[2\nabla_\mu-iB_\mu-\frac{g}{\sqrt{2}}W\gamma_\mu-i\frac{1}{2\sqrt{2}}H_{\nu\rho}\gamma^{\nu\rho}\gamma_\mu\right]\epsilon\,=\,0\,,
\end{equation}
where $\epsilon$ is a complex Dirac spinor and we define
\begin{align} \label{hhbbdef}
H_{\mu\nu}\,\equiv&\,\overline{F}_{\mu\nu}^{78}\,=\,\frac{1}{2}\left(e^{-\lambda_1-\lambda_2-\lambda_3}F_{\mu\nu}^{0}+e^{-\lambda_1+\lambda_2+\lambda_3}F_{\mu\nu}^{1}+e^{\lambda_1-\lambda_2+\lambda_3}F_{\mu\nu}^{2}+e^{\lambda_1+\lambda_2-\lambda_3}F_{\mu\nu}^{3}\right)\,, \notag \\
B_\mu\,\equiv&\,-g\left(A^0_\mu+A^1_\mu+A^2_\mu+A^3_\mu\right)-\frac{1}{2}\left(\cosh\left(2\chi\right)-1\right)D_\mu\psi\,.
\end{align}
From the spin-1/2 field variations, we obtain
\begin{align}
\left[\gamma^\mu\partial_\mu\lambda_1+\frac{g}{\sqrt{2}}\partial_{\lambda_1}W+i\frac{1}{2\sqrt{2}}\gamma^{\mu\nu}\overline{F}_{\mu\nu}^{12}\right]\epsilon\,=&\,0\,, \notag \\
\left[\gamma^\mu\partial_\mu\lambda_2+\frac{g}{\sqrt{2}}\partial_{\lambda_2}W+i\frac{1}{2\sqrt{2}}\gamma^{\mu\nu}\overline{F}_{\mu\nu}^{34}\right]\epsilon\,=&\,0\,, \notag \\
\left[\gamma^\mu\partial_\mu\lambda_3+\frac{g}{\sqrt{2}}\partial_{\lambda_3}W+i\frac{1}{2\sqrt{2}}\gamma^{\mu\nu}\overline{F}_{\mu\nu}^{56}\right]\epsilon\,=&\,0\,, \notag \\
\left[\gamma^\mu\partial_\mu\chi+\frac{g}{\sqrt{2}}\partial_\chi{W}+i\frac{1}{2}\partial_\chi{B}_\mu\gamma^\mu\right]\epsilon\,=&\,0\,.
\end{align}

The scalar potential has the maximally supersymmetric $SO(8)$-invariant $AdS_4$ vacuum with all scalar fields vanishing and $\mathcal{P}_*\,=\,-6$. The vacuum uplifts to the $AdS_4\times{S}^7$ solution of eleven-dimensional supergravity and is dual to the ABJM theory, \cite{Aharony:2008ug}. There is another $\mathcal{N}=2$ supersymmetric $SU(3)\times{U}(1)_R$-invariant $AdS_4$ vacuum known as the Warner fixed point, \cite{Warner:1983vz, Warner:1983du},
\begin{equation} \label{fixedp}
\tanh\chi\,=\,\frac{1}{\sqrt{3}}\,, \qquad \psi\,=\,\frac{\pi}{2}\,, \qquad \tanh\lambda_i\,=\,2-\sqrt{3}\,, \qquad \varphi_i\,=\,\pi\,,
\end{equation}
with $\mathcal{P}_*\,=\,-\frac{9\sqrt{3}}{2}$. The vacuum uplifts to eleven-dimensional supergravity, \cite{Corrado:2001nv}, and is dual to the mass-deformed ABJM theory, \cite{Benna:2008zy, Klebanov:2008vq}. There is a holographic RG flow from the $SO(8)$ vacuum to the $SU(3)\times{U(1)}$ vacuum, \cite{Ahn:2000aq, Ahn:2000mf, Bobev:2009ms}. The radius of the $AdS_4$ is given by
\begin{equation} \label{radii}
L_{AdS_4}^2\,=\,-\frac{3}{g^2\mathcal{P}_*}\,=\,\left\{ \begin{aligned} &\frac{1}{2g^2}\,, \qquad \,\,\,\,\,\,\, SO(8)\,, \\ &\frac{2}{3\sqrt{3}g^2}\,, \qquad SU(3)\times{U}(1)_R\,, \end{aligned} \right.
\end{equation}
Further details of the vacua in this truncation can be found in \cite{Bobev:2010ib}.

From the AdS/CFT correspondence, the free energy of pure $AdS_4$ with an asymptotic boundary of $S^3$ is given by $e.g.$, \cite{Bobev:2018uxk},
\begin{equation} \label{flads4}
\mathcal{F}_{S^3}\,=\,\frac{\pi{L}^2_{AdS_4}}{2G_N^{(4)}}\,,
\end{equation}
where $G_4^{(4)}$ is the four-dimensional Newton's constant. This matches with the field theory free energy of the ABJM theory and the mass-deformed ABJM theory at large $N$, respectively,
\begin{align} \label{fabjm}
\mathcal{F}_{S^3}^{\text{ABJM}}\,=&\,\frac{\sqrt{2}\pi}{3}N^{3/2}\,, \notag \\
\mathcal{F}_{S^3}^{\text{mABJM}}\,=&\,\frac{4\sqrt{2}\pi}{9\sqrt{3}}N^{3/2}\,.
\end{align}
Furthermore, the ratio below is universal, \cite{Jafferis:2011zi},
\begin{equation} \label{lfratio}
\frac{\mathcal{F}_{S^3}}{L_{AdS_4}^2}\,=\,\frac{2\sqrt{2}\pi}{3}g^2N^{3/2}\,.
\end{equation}

When the complex scalar field, $(\chi,\psi)$, is vanishing, the truncation reduces to the STU model with three scalar fields and four gauge fields. By setting all gauge fields equal and all scalar fields vanish, the truncation further reduces to minimal gauged supergravity. See appendix \ref{appA} for details.

\section{$AdS_2$ ansatz} \label{sec3}

We consider the background with the gauge fields,
\begin{align}
ds^2\,=&\,e^{2V}ds_{AdS_2}^2+f^2dy^2+h^2dz^2\,, \notag \\
A^\alpha\,=&\,a^\alpha{d}z\,,
\end{align}
where $ds_{AdS_2}^2$ is a unit radius metric on $AdS_2$ and $V$, $f$, $h$, and $a^\alpha$, $\alpha\,=\,0,\,\ldots,\,3$, are functions of the coordinate $y$ only. The scalar fields, $\chi$ and $\lambda_i$, $i\,=\,1,\ldots,3$, are also functions of the coordinate $y$. In order to avoid partial differential equations from the equations of motion for gauge fields, we take the scalar field, $\psi$, to be $\psi\,=\,\bar{\psi}z$ where $\bar{\psi}$ is a constant. This brings
\begin{equation}
B_\mu{d}x^\mu\,\equiv\,B_zdz\,,
\end{equation}
where $B_z$ is a function of the coordinate $y$.

We employ an orthonormal frame,
\begin{equation}
e^a\,=\,e^V\bar{e}^a\,, \qquad e^2\,=\,fdy\,, \qquad e^3\,=\,hdz\,,
\end{equation}
where $\bar{e}^a$ is an orthonormal frame for $ds_{AdS_2}^2$. The frame components of the field strengths are given by
\begin{equation}
F_{23}^\alpha\,=\,f^{-1}h^{-1}\left(a^\alpha\right)'\,.
\end{equation}

The equations of motion for gauge fields are combined and integrated to give the integrals of motion,
\begin{align} \label{constantmo1}
e^{2V}\left(e^{-2\lambda_1+2\lambda_2+2\lambda_3}F_{23}^1-e^{2\lambda_1-2\lambda_2+2\lambda_3}F_{23}^2\right)\,=&\,\mathcal{E}_{F_1}\,, \notag \\
e^{2V}\left(e^{2\lambda_1-2\lambda_2+2\lambda_3}F_{23}^2-e^{2\lambda_1+2\lambda_2-2\lambda_3}F_{23}^3\right)\,=&\,\mathcal{E}_{F_2}\,, \notag \\
e^{2V}\left(e^{2\lambda_1+2\lambda_2-2\lambda_3}F_{23}^3-e^{-2\lambda_1+2\lambda_2+2\lambda_3}F_{23}^1\right)\,=&\,\mathcal{E}_{F_3}\,, \notag \\
\end{align}
and
\begin{align} \label{constantmo2}
e^{2V}\left(e^{-2\lambda_1-2\lambda_2-2\lambda_3}F_{23}^0+e^{-2\lambda_1+2\lambda_2+2\lambda_3}F_{23}^1\right)\,=&\,\mathcal{E}_{R_1}\,, \notag \\
e^{2V}\left(e^{-2\lambda_1-2\lambda_2-2\lambda_3}F_{23}^0+e^{2\lambda_1-2\lambda_2+2\lambda_3}F_{23}^2\right)\,=&\,\mathcal{E}_{R_2}\,, \notag \\
e^{2V}\left(e^{-2\lambda_1-2\lambda_2-2\lambda_3}F_{23}^0+e^{2\lambda_1+2\lambda_2-2\lambda_3}F_{23}^3\right)\,=&\,\mathcal{E}_{R_3}\,,
\end{align}
with
\begin{equation}
\left(e^{2V+2\lambda_1+2\lambda_2-2\lambda_3}F_{23}^3\right)'\,=\,-e^{2V}fh^{-1}\frac{1}{2}\sinh^2\left(2\chi\right)D_z\psi\,,
\end{equation}
where $\mathcal{E}_{F_i}$ and $\mathcal{E}_{R_i}$ are constant. Among the six integrals of motion in \eqref{constantmo1} and \eqref{constantmo2}, only three of them are independent, $e.g.$, three in \eqref{constantmo2}, and others can be obtained by combining the three independent ones.

\subsection{BPS equations}

We employ the gamma matrices,
\begin{equation}
\gamma^m\,=\,\Gamma^m\otimes\sigma^3\,, \qquad \gamma^2\,=\,\mathbb{I}_2\otimes\sigma^1\,, \qquad \gamma^3\,=\,\mathbb{I}_2\otimes\sigma^2\,,
\end{equation}
where $\Gamma^m$ are two-dimensional gamma matrices of mostly plus signature. The spinors are given by
\begin{equation}
\epsilon\,=\,\psi\otimes\chi\,,
\end{equation}
and the two-dimensional spinor on $AdS_2$ satisfies
\begin{equation}
D_m\psi\,=\,\frac{1}{2}\kappa\Gamma_m\psi\,,
\end{equation}
where $\kappa\,=\,\pm1$ fixes the chirality. The spinor, $\chi$, is given by 
\begin{equation}
\chi\,=\,e^{V/2}e^{isz/2}\left(
\begin{array}{ll}
 \sin\frac{\xi}{2} \\
 \cos\frac{\xi}{2}
\end{array}
\right)\,,
\end{equation}
where the constant, $s$, is the gauge dependent charge under the action of azimuthal Killing vector, $\partial_z$.

We consider the case of $\sin\xi\ne0$. The complete BPS equations are derived in appendix \ref{appB} and are given by
\begin{align} \label{bpsbps}
f^{-1}\xi'\,=&\,\sqrt{2}gW\cos\xi+\kappa{e}^{-V}\,, \notag \\
f^{-1}V'\,=&\,\frac{g}{\sqrt{2}}W\sin\xi\,, \notag \\
f^{-1}\lambda_i'\,=&\,-\frac{g}{\sqrt{2}}\partial_{\lambda_i}W\sin\xi\,, \notag \\
f^{-1}\chi'\,=&\,-\frac{g}{\sqrt{2}}\frac{\partial_\chi{W}}{\sin\xi}\,, \notag \\
f^{-1}\frac{h'}{h}\,=&\,\frac{1}{\sin\xi}\left(\kappa{e}^{-V}\cos\xi+\frac{gW}{\sqrt{2}}\left(1+\cos^2\xi\right)\right)\,, 
\end{align}
with two constraints,
\begin{align} \label{constraintconstraint}
\left(s-B_z\right)\sin\xi\,=&\,-\sqrt{2}gWh\cos\xi-\kappa{h}e^{-V}\,, \notag \\
\sqrt{2}g\partial_\chi{W}\cos\xi\,=&\,\partial_\chi{B}_z\sin\xi{h}^{-1}\,.
\end{align}
The scalar-field dressed field strengths are given by
\begin{align} \label{fstfst}
\overline{F}_{23}^{12}\,=&\,-g\partial_{\lambda_1}W\cos\xi\,, \notag \\
\overline{F}_{23}^{34}\,=&\,-g\partial_{\lambda_2}W\cos\xi\,, \notag \\
\overline{F}_{23}^{56}\,=&\,-g\partial_{\lambda_3}W\cos\xi\,, \notag \\
H_{23}\,=&\,-gW\cos\xi-\sqrt{2}\kappa{e}^{-V}\,.
\end{align}
We have checked that the BPS equations solve the equations of motion from the Lagrangian in \eqref{mlag} as presented in appendix \ref{appA}.

\subsection{Integrals of motion}

Following the observation made in \cite{Arav:2022lzo}, we find an integral of the BPS equations,
\begin{equation} \label{hevks}
he^{-V}\,=\,k\sin\xi\,,
\end{equation}
where $k$ is a constant. Hence, at the poles of the spindle solution, $h=0$, we also have $\sin\xi=0$. From \eqref{bpsbps} and \eqref{constraintconstraint}, we obtain
\begin{equation} \label{xipxip}
\xi'\,=\,-k^{-1}\left(s-B_z\right)\left(e^{-V}f\right)\,,
\end{equation}
and then the two constraints in \eqref{constraintconstraint} can be written as
\begin{align} \label{constrainttwo}
\left(s-B_z\right)\,=&\,-k\left[\sqrt{2}gWe^V\cos\xi+\kappa\right]\,, \notag \\
\sqrt{2}g\partial_\chi{W}\cos\xi\,=&\,k^{-1}e^{-V}\partial_\chi{B}_z\,.
\end{align}

By employing the field strengths in \eqref{fstfst}, we can express the integrals of motion by
\begin{align} \label{er123}
\mathcal{E}_{R_1}\,=&\,e^V\left[2ge^V\cos\xi-\sqrt{2}\kappa{e}^{-\lambda_1}\cosh\left(\lambda_2+\lambda_3\right)\right]\,, \notag \\
\mathcal{E}_{R_2}\,=&\,e^V\left[2ge^V\cos\xi-\sqrt{2}\kappa{e}^{-\lambda_2}\cosh\left(\lambda_3+\lambda_1\right)\right]\,, \notag \\
\mathcal{E}_{R_3}\,=&\,e^V\left[2ge^V\cos\xi-\sqrt{2}\kappa{e}^{-\lambda_3}\cosh\left(\lambda_1+\lambda_2\right)\right]\,,
\end{align}
and
\begin{align}
\mathcal{E}_{F_1}\,=&\,\sqrt{2}\kappa{e}^Ve^{\lambda_3}\sinh\left(\lambda_1-\lambda_2\right)\,, \notag \\
\mathcal{E}_{F_2}\,=&\,\sqrt{2}\kappa{e}^Ve^{\lambda_1}\sinh\left(\lambda_2-\lambda_3\right)\,, \notag \\
\mathcal{E}_{F_3}\,=&\,\sqrt{2}\kappa{e}^Ve^{\lambda_2}\sinh\left(\lambda_3-\lambda_1\right)\,.
\end{align}

\subsection{Boundary conditions for spindle solutions} \label{sec33}

We fix the metric to be in conformal gauge,
\begin{equation}
f\,=\,e^V\,,
\end{equation}
and the metric is given by
\begin{equation}
ds^2\,=\,e^{2V}\left[ds_{AdS_2}^2+ds_{\mathbbl{\Sigma}}^2\right]\,,
\end{equation}
where we have
\begin{equation}
ds_{\mathbbl{\Sigma}}^2\,=\,dy^2+k^2\sin^2\xi{d}z^2\,,
\end{equation}
for the metric on a spindle, ${\mathbbl{\Sigma}}$. For the spindle solutions, there are two poles at $y=y_{N,S}$ with deficit angles of $2\pi\left(1-\frac{1}{n_{N,S}}\right)$. The azimuthal angle, $z$, has a period which we set
\begin{equation}
\Delta{z}\,=\,2\pi\,.
\end{equation}

\subsubsection{Analysis of the BPS equations}

Following the argument in section 3.3.1 of \cite{Arav:2022lzo} we study the BPS equations to find the spindle solutions. At the poles of the spindle solution, $y\,=\,y_{N,S}$, as we have $k\sin\xi\rightarrow0$, we find $\cos\xi\rightarrow\pm1$ if $k\ne0$. Thus, we express $\cos\xi_{N,S}\,=\,(-1)^{t_{N,S}}$ with $t_{N,S}\in\{0,1\}$. We choose $y_N<y_S$ and $y\in[y_N,y_S]$. We assume that the deficit angles at the poles are $2\pi\left(1-\frac{1}{n_{N,S}}\right)$ with $n_{N,S}\ge1$. Then we require the metric to have $|\left(k\sin\xi\right)'|_{N,S}\,=\,\frac{1}{n_{N,S}}$. From the symmetry of BPS equations in \eqref{hsymm} and \eqref{hevks}, we further choose
\begin{equation}
h\ge0\,, \qquad \Leftrightarrow \qquad k\sin\xi\ge0\,.
\end{equation}
Then we find $\left(k\sin\xi\right)'|_N>0$ and $\left(k\sin\xi\right)'|_S<0$. Hence, we impose 
\begin{equation}
\left(k\sin\xi\right)'|_{N,S}\,=\,\frac{(-1)^{l_{N,S}}}{n_{N,S}}\,, \qquad l_N\,=\,0\,,l_S\,=\,1\,.
\end{equation}

Two different classes of spindle solutions, the twist and the anti-twist, are known, \cite{Ferrero:2021etw}. Spinors have the same chirality at the poles in the twist solutions and opposite chiralities in the anti-twist solutions as
\begin{align} \label{cosxi1}
\cos\xi|_{N,S}\,=\,(-1)^{t_{N,S}}; \qquad &\text{Twist:} \,\, \qquad \quad \left(t_N,t_S\right)\,=\,\left(1,1\right) \quad \text{or} \quad \left(0,0\right)\,, \notag \\
&\text{Anti-Twist:} \quad \left(t_N,t_S\right)\,=\,\left(1,0\right) \quad \text{or} \quad \left(0,1\right)\,.
\end{align}

As we find $\left(k\sin\xi\right)'\,=\,-\cos\xi\left(s-B_z\right)$ from the BPS equation in \eqref{xipxip}, we obtain
\begin{equation}
\left(s-B_z\right)|_{N,S}\,=\,\frac{1}{n_{N,S}}(-1)^{l_{N,S}+t_{N,S}+1}\,.
\end{equation}
We consider the flux quantization for R-symmetry flux. From \eqref{hhbbdef} we have $F^R\,=\,dB+d\left(\frac{1}{2}\left(\cosh\left(2\chi\right)-1\right)D\psi\right)$. At the poles, as $\chi=0$ unless $D\psi=0$, the second term on the right hand side of $F^R$ does not contribute to the flux quantization. Then, we find the R-symmetry flux quantized to be
\begin{equation} \label{rsymmq}
\frac{1}{2\pi}\int_{\mathbbl{\Sigma}}F^R\,\equiv\,\frac{1}{2\pi}\int_{\mathbbl{\Sigma}}-g\left(F^0+F^1+F^2+F^3\right)\,=\,\frac{n_N(-1)^{t_S+1}+n_S(-1)^{t_N+1}}{n_Nn_S}\,.
\end{equation}

We have $\partial_zB=-\frac{1}{2}\sinh\left(2\chi\right)D_z\psi$. Once again, as $\chi=0$ unless $D\psi=0$ at the poles, we find $\partial_\chi{B}_z=0$ at the poles.  From the constraint in \eqref{constrainttwo} we also find $\partial_\chi{W}=0$ at the poles. Hence, we have
\begin{equation} \label{dbdw}
\partial_\chi{B}_z|_{N,S}\,=\,\partial_\chi{W}|_{N,S}\,=\,0\,.
\end{equation}

We further assume that the complex scalar field, $(\chi,\psi)$, is non-vanishing at the poles and we find
\begin{equation} \label{chitopsi}
\chi|_N\,,\chi|_S\,\ne\,0\,, \qquad \Rightarrow \qquad D_z\psi|_N\,=\,D_z\psi|_S\,=\,0\,.
\end{equation}
Thus, we find that the flux charging the complex scalar field should vanish,
\begin{equation} \label{msymmq}
\frac{1}{2\pi}\int_{\mathbbl{\Sigma}}g\left(F^0-F^1-F^2-F^3\right)\,=\,\left(D_z\psi\right)|_{y_N}^{y_S}\,=\,0\,.
\end{equation}

Note that, from \eqref{chitopsi} and the second condition in \eqref{dbdw}, we find
\begin{equation} \label{dw=0}
\left(e^{2\lambda_1}+e^{2\lambda_2}+e^{2\lambda_3}-e^{2\lambda_1+2\lambda_2+2\lambda_3}\right)|_{N,S}\,=\,0\,, \quad \Rightarrow \quad W|_{N,S}\,=\,-e^{\lambda_1+\lambda_2+\lambda_3}|_{N,S}\,.
\end{equation}
In order to find the values of the integrals of motion, $\mathcal{E}_{R_i}$ in \eqref{er123}, we introduce two quantities,
\begin{equation} \label{defm1m2}
M_{(1)}\,\equiv\,ge^{\lambda_1+\lambda_2+\lambda_3}e^V\,, \qquad M_{(2)}\,\equiv\,-\frac{\kappa}{\sqrt{2}}M_{(1)}+2M_{(1)}^2\cos\xi\,,
\end{equation}
and note that $M_{(1)}>0$. The integrals of motion, \eqref{er123}, are functions of $V$, $\lambda_i$ and $\cos\xi$. We can eliminate $V$ by using the first equation in \eqref{constrainttwo} and $\cos\xi$ by \eqref{cosxi1}. Then we find the integrals of motion to be
\begin{equation}
\mathcal{E}_{R_i}\,=\,-\frac{\kappa}{\sqrt{2}g}M_{(1)}e^{-2\lambda_i}+\frac{1}{g}M_{(2)}e^{-2\lambda_1-2\lambda_2-2\lambda_3}\,,
\end{equation}
with
\begin{align}
M_{(1)}|_{N,S}\,=&\,\frac{1}{\sqrt{2}}(-1)^{t_{N,S}}\kappa-\frac{1}{\sqrt{2}kn_{N,S}}(-1)^{l_{N,S}}\,, \notag \\
M_{(2)}|_{N,S}\,=&\,\frac{1}{2}(-1)^{t_{N,S}}+\frac{1}{k^2n_{N,S}^2}(-1)^{t_{N,S}}-\frac{3\kappa}{2kn_{N,S}}(-1)^{l_{N,S}}\,.
\end{align}
Finally, we can eliminate one of the scalar fields, $\lambda_i$, say $\lambda_3$, by the condition on the left hand side of \eqref{dw=0}. Hence, we have three independent integrals of motion in terms of two scalar fields, $\lambda_1$ and $\lambda_2$. As the integrals of motion have identical values at the poles, we find three algebraic equations with four unknowns, $(\lambda_{1N},\,\lambda_{1S} \,,\lambda_{2N},\,\lambda_{2S})$,
\begin{align} \label{assoeq}
\mathcal{E}_{R_1}(\lambda_{1N},\lambda_{2N})\,&=\,\mathcal{E}_{R_1}(\lambda_{1S},\lambda_{2S})\,, \notag \\
\mathcal{E}_{R_2}(\lambda_{1N},\lambda_{2N})\,&=\,\mathcal{E}_{R_2}(\lambda_{1S},\lambda_{2S})\,, \notag \\
\mathcal{E}_{R_3}(\lambda_{1N},\lambda_{2N})\,&=\,\mathcal{E}_{R_3}(\lambda_{1S},\lambda_{2S})\,.
\end{align}
Unlike the case for the Leigh-Strassler theory in \cite{Arav:2022lzo}, we are one equation short to solve for the values of all scalar fields at the poles.{\footnote {In \cite{Arav:2022lzo}, in the five-dimensional supergravity model, three $U(1)$ gauge fields, $F^I$, $I=1,2,3$, produce two conserved quantities, $\mathcal{E}_R(V,\alpha,\beta)$ and $\mathcal{E}_F(V,\alpha,\beta)$, where $V$ is again the warp factor and $\alpha$ and $\beta$ are the scalar fields. One can fix $V$ from the flux quantization of $\left(s-Q_z\right)$ and fix $\beta$ from $\partial{W}=0$. Then $\mathcal{E}_R(\alpha_S)=\mathcal{E}_R(\alpha_N)$ and $\mathcal{E}_F(\alpha_S)=\mathcal{E}_F(\alpha_N)$ fix the values of the scalar field at the poles, $(\alpha_S,\alpha_N)$. The numbers of unknowns and the numbers of constraints from the conserved quantities match. The values of all the other fields at the poles are determined subsequently.}} We need to find an additional constraint to determine all the values of the fields at the poles and it will be found in the analysis of fluxes in the next subsection. 

\subsubsection{Fluxes}

In appendix \ref{appB} we have obtained the expressions of field strengths in terms of the scalar fields, warp factors, the angle, $\xi$, and $k$,
\begin{equation}
F^\alpha_{yz}\,=\,\left(a^\alpha\right)'\,=\,\left(\mathcal{I}^{(\alpha)}\right)'\,,
\end{equation}
where we have
\begin{align}
\mathcal{I}^{(0)}\,\equiv&\,\frac{1}{\sqrt{2}}ke^V\cos\xi\,e^{\lambda_1+\lambda_2+\lambda_3}\,, \notag \\
\mathcal{I}^{(1)}\,\equiv&\,\frac{1}{\sqrt{2}}ke^V\cos\xi\,e^{\lambda_1-\lambda_2-\lambda_3}\,, \notag \\
\mathcal{I}^{(2)}\,\equiv&\,\frac{1}{\sqrt{2}}ke^V\cos\xi\,e^{-\lambda_1+\lambda_2-\lambda_3}\,, \notag \\
\mathcal{I}^{(3)}\,\equiv&\,\frac{1}{\sqrt{2}}ke^V\cos\xi\,e^{-\lambda_1-\lambda_2+\lambda_3}\,.
\end{align}
Thus we find that the fluxes are solely given by the data at the poles,
\begin{equation}
\frac{p_\alpha}{n_Nn_S}\,\equiv\,\frac{1}{2\pi}\int_{\mathbbl{\Sigma}}gF^\alpha\,=\,g\mathcal{I}^\alpha|_N^S\,.
\end{equation}

From \eqref{fourvectors} we define R-symmetry and two flavor symmetry fluxes by, respectively,
\begin{align} \label{i04}
\mathcal{I}_0|_{N,S}\,\equiv&\,-\left(\mathcal{I}^{(0)}+\mathcal{I}^{(1)}+\mathcal{I}^{(2)}+\mathcal{I}^{(3)}\right)|_{N,S}\,, \notag \\
\mathcal{I}_{\Delta_1}|_{N,S}\,\equiv&\,\frac{1}{2}\left(\mathcal{I}^{(1)}-\mathcal{I}^{(2)}\right)|_{N,S}\,, \notag \\
\mathcal{I}_{\Delta_2}|_{N,S}\,\equiv&\,\frac{1}{\sqrt{3}}\left(\mathcal{I}^{(1)}+\mathcal{I}^{(2)}-2\mathcal{I}^{(3)}\right)|_{N,S}\,.
\end{align}
By \eqref{i04} we find
\begin{align}
g\mathcal{I}_0|_{N,S}\,=&\,-g\left(\mathcal{I}^{(0)}+\mathcal{I}^{(1)}+\mathcal{I}^{(2)}+\mathcal{I}^{(3)}\right)|_{N,S} \notag \\
=&\,-\frac{1}{\sqrt{2}}gke^V\cos\xi\left(e^{\lambda_1+\lambda_2+\lambda_3}+e^{\lambda_1-\lambda_2-\lambda_3}+e^{-\lambda_1+\lambda_2-\lambda_3}+e^{-\lambda_1-\lambda_2+\lambda_3}\right)|_{N,S} \notag \\
=&\,-\frac{1}{\sqrt{2}}gke^V\cos\xi\,e^{-\lambda_1-\lambda_2-\lambda_3}\left(e^{2\lambda_1+2\lambda_2+2\lambda_3}+e^{2\lambda_1}+e^{2\lambda_2}+e^{2\lambda_3}\right)|_{N,S} \notag \\
=&\,-\frac{1}{\sqrt{2}}gke^V\cos\xi2e^{\lambda_1+\lambda_2+\lambda_3}|_{N,S} \notag \\
=&\,\sqrt{2}kM_{(1)}|_{N,S}(-1)^{t_{N,S}+1}\,,
\end{align}
and
\begin{align}
&g\left(\mathcal{I}^{(0)}-\mathcal{I}^{(1)}-\mathcal{I}^{(2)}-\mathcal{I}^{(3)}\right)|_{N,S} \notag \\
=&\,\frac{1}{\sqrt{2}}gke^V\cos\xi\left(e^{\lambda_1+\lambda_2+\lambda_3}-e^{\lambda_1-\lambda_2-\lambda_3}-e^{-\lambda_1+\lambda_2-\lambda_3}-e^{-\lambda_1-\lambda_2+\lambda_3}\right)|_{N,S} \notag \\
=&\,\frac{1}{\sqrt{2}}gke^V\cos\xi\,e^{-\lambda_1-\lambda_2-\lambda_3}\left(-e^{2\lambda_1+2\lambda_2+2\lambda_3}+e^{2\lambda_1}+e^{2\lambda_2}+e^{2\lambda_3}\right)|_{N,S}\,=\,0\,,
\end{align}
where we used \eqref{dw=0}. Thus we recover the R-symmetry flux quantization, \eqref{rsymmq}, and vanishing of the flux of massive vector field, \eqref{msymmq}, respectively,
\begin{align}
g\mathcal{I}_0|_N^S\,=\,-g\left(\mathcal{I}^{(0)}+\mathcal{I}^{(1)}+\mathcal{I}^{(2)}+\mathcal{I}^{(3)}\right)|_N^S\,=&\,\frac{n_N(-1)^{t_S+1}+n_S(-1)^{t_N+1}}{n_Nn_S}\,, \notag \\
g\left(\mathcal{I}^{(0)}-\mathcal{I}^{(1)}-\mathcal{I}^{(2)}-\mathcal{I}^{(3)}\right)|_N^S\,=&\,0\,.
\end{align}
Furthermore, we define the fluxes of two flavor symmetry vectors by
\begin{align} \label{fsymmq1}
\frac{p_{F_1}}{n_Nn_S}\,\equiv&\,g\mathcal{I}_{\Delta_1}|_N^S\,=\,\frac{1}{2}g\left(\mathcal{I}^{(1)}-\mathcal{I}^{(2)}\right)|_N^S \notag \\
=&\,\frac{1}{2\sqrt{2}}gke^V\cos\xi\left(e^{\lambda_1-\lambda_2-\lambda_3}-e^{-\lambda_1+\lambda_2-\lambda_3}\right)|_N^S \notag \\
=&\,\frac{1}{2\sqrt{2}}kM_{(1)}(-1)^{t_{N,S}}\left(e^{-2\lambda_2-2\lambda_3}-e^{-2\lambda_3-2\lambda_1}\right)|_N^S\,, \\ \notag \\ \label{fsymmq2}
\frac{p_{F_2}}{n_Nn_S}\,\equiv&\,g\mathcal{I}_{\Delta_2}|_N^S\,=\,\frac{1}{\sqrt{3}}g\left(\mathcal{I}^{(1)}+\mathcal{I}^{(2)}-2\mathcal{I}^{(3)}\right)|_N^S \notag \\
=&\,\frac{1}{\sqrt{6}}gke^V\cos\xi\left(e^{\lambda_1-\lambda_2-\lambda_3}+e^{-\lambda_1+\lambda_2-\lambda_3}-2e^{-\lambda_1-\lambda_2+\lambda_3}\right)|_N^S \notag \\
=&\,\frac{1}{\sqrt{6}}kM_{(1)}(-1)^{t_{N,S}}\left(e^{-2\lambda_2-2\lambda_3}+e^{-2\lambda_3-2\lambda_1}-2e^{-2\lambda_1-2\lambda_2}\right)|_N^S\,,
\end{align}
where $p_{F_1}$ and $p_{F_2}$ are integers. From these two constraints, we should be able to find expressions of $k$ and one of the scalar fields, $\lambda_i$, which was not determined in \eqref{assoeq}.{\footnote{In the analysis of 4d Leigh-Strassler theory in \cite{Arav:2022lzo}, as there is one flavor symmetry vector, there is only one constraint from the flavor flux. Unlike \cite{Arav:2022lzo}, we have two flavor symmetry fluxes and this is where we find the additional constraint which was missing in the analysis of \eqref{assoeq}.}} 

{\bf Summary of the constraints to determine all the boundary conditions:} Let us summarize the constraints we have obtained to determine all the boundary conditions. By solving seven associated equations, the left hand side of \eqref{dw=0}, \eqref{assoeq}, \eqref{fsymmq1}, and \eqref{fsymmq2}, we can determine the values of the scalar fields, $\lambda_1$, $\lambda_2$, $\lambda_3$, at the north and south poles and also the value of the constant, $k$, in terms of $n_{N,S}$, $t_{N,S}$, $p_{F_1}$, and $p_{F_2}$. Then the values of $V$ at the poles are determined from the definition of $M_{(1)}$ in \eqref{defm1m2}. This fixes all the boundary conditions except the hyper scalar field, $\chi$, which will be freely chosen when constructing the solutions explicitly. However, the constraint equations are quite complicated and it appears to be not easy to solve them.

Even though we are not able to solve for the boundary conditions in terms of $n_{N,S}$, $t_{N,S}$, $p_{F_1}$, and $p_{F_2}$ analytically, if we choose numerical values of $n_{N,S}$, $t_{N,S}$, $p_{F_1}$, and $p_{F_2}$, the constraints can be solved to determine all the boundary conditions. For instance, in the anti-twist class, for the choice of
\begin{align} \label{ninput}
n_N\,=&\,8\,, \qquad n_S\,=\,1\,, \qquad p_{F_1}\,=\,1\,, \qquad p_{F_2}\,=\,2\,, \notag \\
g\,=&\,1\,, \qquad \kappa\,=\,-1\,,
\end{align}
we find the boundary conditions to be
\begin{align} \label{noutput}
e^{-2\lambda_{1N}}\,\approx&\,0.0452746\,, \qquad \,\,\,\,\,\, e^{-2\lambda_{1S}}\,\approx\,0.012202\,, \notag\\
e^{-2\lambda_{2N}}\,\approx&\,0.110402\,, \qquad \quad \,\,\, e^{-2\lambda_{2S}}\,\approx\,0.101903\,, \notag\\
e^{-2\lambda_{3N}}\,\approx&\,6.39148\,, \qquad \qquad e^{-2\lambda_{3S}}\,\approx\,8.75297\,, \notag\\
k\,\approx&\,0.631939\,.
\end{align}
In this way, without finding analytic expression of the Bekenstein-Hawking entropy, we can determine numerical value for each choice of $n_{N,S}$, $t_{N,S}$, $p_{F_1}$, and $p_{F_2}$. Furthermore, we will be able to construct the solutions explicitly numerically.

\subsubsection{The Bekenstein-Hawking entropy} \label{sec333}

The $AdS_2\times\mathbbl{\Sigma}$ solution would be the horizon of a presumed black hole which asymptotes to the $\mathcal{N}=2$ $AdS_4$ vacuum dual to the mass-deformed ABJM theory. We calculate the Bekenstein-Hawking entropy of the presumed black hole solution.

From the AdS/CFT dictionary, \eqref{flads4} and \eqref{lfratio}, for the four-dimensional Newton's constant, we have,
\begin{equation}
\frac{1}{2G_N^{(4)}}\,=\,\frac{2\sqrt{2}}{3}g^2N^{3/2}\,.
\end{equation}
Then the two-dimensional Newton's constant is given by
\begin{equation}
\left(G_N^{(2)}\right)^{-1}\,=\,\left(G_N^{(4)}\right)^{-1}\Delta{z}\int_{y_N}^{y_S}|fh|dy\,.
\end{equation}
Employing the BPS equations, we find
\begin{equation}
fh\,=\,ke^Vf\sin\xi\,=\,-\frac{k}{\kappa}\left(e^{2V}\cos\xi\right)'\,.
\end{equation}
Hence, the Bekenstein-hawking entropy is expressed by the data at the poles,
\begin{align} \label{bhent}
S_{\text{BH}}\,=&\,\frac{1}{4G_N^{(2)}}\,=\,\frac{2\sqrt{2}\pi}{3}N^{3/2}g^2\left(-\frac{k}{\kappa}\right)\left[e^{2V}\cos\xi\right]_N^S \notag \\
=&\,-\frac{2\sqrt{2}N^{3/2}k}{3\kappa}\left(M_{(1)}^2|_S\,e^{-2\left(\lambda_{1S}+\lambda_{2S}+\lambda_{3S}\right)}(-1)^{t_S}-M_{(1)}^2|_N\,e^{-2\left(\lambda_{1N}+\lambda_{2N}+\lambda_{3N}\right)}(-1)^{t_N}\right)\,,
\end{align}
where we expressed the Bekenstein-Hawking entropy in terms of $M_{(1)}$.

As we can determine the numerical values of the boundary conditions for each choice of $n_{N,S}$, $t_{N,S}$, $p_{F_1}$, and $p_{F_2}$, we can find the numerical value of the Bekenstein-Hawking entropy as well. For instance, for the choice of \eqref{ninput}, the Bekenstein-Hawking entropy is given by $S_{\text{BH}}\,\approx\,0.0226955N^{3/2}$.

Furthermore, when there is no flavor charges, $p_{F_1}=p_{F_2}=0$, we perform a non-trivial check that the numerical value of the Bekenstein-Hawking entropy precisely matches the value obtained from the formula given in \eqref{minent} for the solutions from minimal gauged supergravity.

\section{Solving the BPS equations} \label{sec4}

\subsection{Analytic solutions for minimal gauged supergravity via W}

In minimal gauged supergravity associated with the Warner $\mathcal{N}=2$ $AdS_4$ vacuum, utilizing the class of $AdS_2\times\mathbbl{\Sigma}$ solutions in \cite{Ferrero:2020twa}, we find solutions in the anti-twist class to the BPS equations in \eqref{bpsbps}, \eqref{constraintconstraint} and \eqref{fstfst}. The scalar fields take the values at the Warner $\mathcal{N}=2$ $AdS_4$ vacuum,
\begin{equation}
\tanh\chi\,=\,\frac{1}{\sqrt{3}}\,, \qquad \psi\,=\,\frac{\pi}{2}\,, \qquad \tanh\lambda_i\,=\,2-\sqrt{3}\,, \qquad \varphi_i\,=\,\pi\,.
\end{equation}
The metric and the gauge field are given by
\begin{align}
ds^2\,=&\,\frac{2}{3\sqrt{3}g^2}\left[\frac{y^2}{4}ds_{AdS_2}^2+\frac{y^2}{q(y)}dy^2+\frac{q(y)}{4y^2}c_0^2dz^2\right]\,, \notag \\
\frac{1}{3}A^0\,=\,A^1\,=\,A^2\,=\,A^3\,=&\,-\left[\frac{2c_0\kappa}{3g}\left(1-\frac{a}{y}\right)+\frac{s}{g}\right]dz\,,
\end{align}
and we have
\begin{equation}
\sin\xi\,=\,-\frac{\sqrt{q(y)}}{y^2}\,, \qquad \cos\xi\,=\,\kappa\frac{2y-a}{y^2}\,.
\end{equation}
Note that for the overall factor in the metric, we have $L_{AdS_4}^2=\frac{2}{3\sqrt{3}g^2}$ for the Warner vacuum from \eqref{radii}. The quartic function is given by
\begin{equation}
q(y)\,=\,y^4-4y^2+4ay-a^2\,,
\end{equation}
and the constants are
\begin{align}
a\,=&\,\frac{n_S^2-n_N^2}{n_S^2+n_N^2}\,, \notag \\
c_0\,=&\,\frac{\sqrt{n_S^2+n_N^2}}{\sqrt{2}n_Sn_N}\,.
\end{align}
We set $n_S>n_N$. For the two middle roots of $q(y)$, $y\in[y_N,y_S]$, we find
\begin{equation}
y_N\,=\,-1+\sqrt{1+a}\,, \qquad y_S\,=\,1-\sqrt{1-a}\,.
\end{equation}
The Bekenstein-Hawking entropy is calculated to give
\begin{align} \label{minent}
S_{\text{BH}}\,=&\,\frac{\sqrt{2}\sqrt{n_S^2+n_N^2}-\left(n_S+n_N\right)}{n_Sn_N}\frac{\pi{L}_{AdS_4}^2}{4G_N^{(4)}} \notag \\
=&\,\frac{\sqrt{2}\sqrt{n_S^2+n_N^2}-\left(n_S+n_N\right)}{n_Sn_N}\frac{1}{2}\mathcal{F}_{S^3}^{\text{mABJM}}\,,
\end{align}
where we employed \eqref{radii} and \eqref{flads4} and $\mathcal{F}_{S^3}^{\text{mABJM}}=\frac{4\sqrt{2}\pi}{9\sqrt{3}}N^{3/2}$, \eqref{fabjm}, is the free energy of mass-deformed ABJM theory.

\subsection{Numerical solutions for $p_{F_1},p_{F_2}\ne0$}

In section \ref{sec33}, although we were not able to find the analytic expressions of the boundary conditions, we were able to determine the numerical values of the boundary conditions for each choice of $n_{N,S}$, $t_{N,S}$, $p_{F_1}$, and $p_{F_2}$. Employing these results for the boundary conditions, we can numerically construct $AdS_2\times\mathbbl{\Sigma}$ solutions in the anti-twist class by solving the BPS equations.{\footnote{As we do not know the analytic expressions of the boundary conditions, we could not exclude the existence of solutions in the twist class. However, we were not able to find any boundary conditions for numerical solutions in the twist class.}}

In order to solve the BPS equations numerically, we start the integration at $y=y_N$ and we choose $y_N=0$. At the poles we have $\sin\xi=0$. We scan over the initial value of $\chi$ at $y=y_N$ in search of a solution for which we have $\sin\xi=0$ in a finite range, $i.e.$, at $y=y_S$. If we find compact spindle solution, our boundary conditions guarantee the fluxes to be properly quantized.

\begin{figure}[t]
\begin{center}
\includegraphics[width=3.2in]{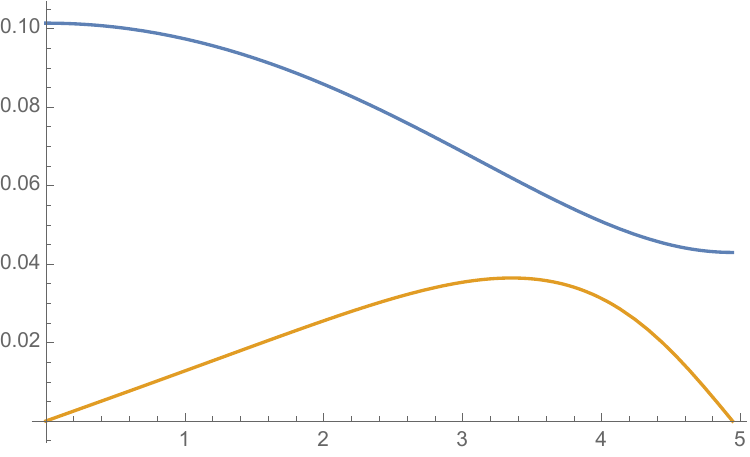} \qquad \includegraphics[width=3.2in]{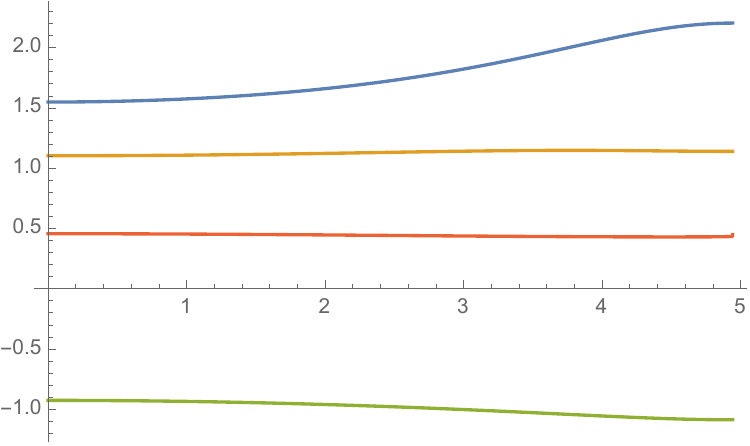}
\caption{{\it A representative $AdS_2\times\mathbbl{\Sigma}$ solution in the anti-twist class for $n_N=8$, $n_S=1$, $p_{F_1}=1$ and $p_{F_2}=2$ in the range of $y=[y_N,y_S]\approx[0,4.9433467]$. The metric functions, $e^V$ (Blue) and $h$ (Orange), are in the left. The scalar fields, $\lambda_1$ (Blue), $\lambda_2$ (Orange), $\lambda_3$ (Green), and $\chi$ (Red) are in the right. Note that $h$ vanishes at the poles.} \label{fig1}}
\end{center}
\end{figure}

We numerically perform the Bekenstein-Hawking entropy integral in section \ref{sec333} and the result matches the Bekenstein-Hawking entropy in \eqref{bhent} with the numerical accuracy of order $10^{-6}$. We present a representative solution in figure \ref{fig1} for the choice in \eqref{ninput} in the range of $y=[y_N,y_S]\approx[0,4.9433467]$. The scalar field, $\chi$, takes the values, $\chi|_N\approx0.455$ and $\chi|_S\approx0.447928$, at the poles. Note that $h$ vanishes at the poles.

There appears to be constraints on the parameter space of $n_{N,S}$, $t_{N,S}$, $p_{F_1}$, and $p_{F_2}$. However, without the analytic expressions of the boundary conditions, it is not easy to specify the constraints.

\section{Conclusions} \label{sec5}

In this paper, we constructed supersymmetric $AdS_2\times\mathbbl{\Sigma}$ solutions which is presumably asymptotic to the Warner $\mathcal{N}=2$ $SU(3)\times{U}(1)_R$ $AdS_4$ vacuum dual to the mass-deformed ABJM theory, where $\mathbbl{\Sigma}$ is a spindle. We derived the complete BPS equations required to construct the solution. We found the flux quantization through the spindle from the analysis of the BPS equations. We found all the necessary algebraic constraints to determine the boundary conditions to construct solutions explicitly. Due to the complexity of the constraints, even though we were not able to solve for the boundary conditions analytically, we were able to determine numerical values of the boundary conditions for each choice of $n_{N,S}$, $t_{N,S}$, $p_{F_1}$, and $p_{F_2}$. Employing this, we explicitly constructed the solutions in the anti-twist class numerically and calculated the numerical values of Bekenstein-Hawking entropy.

As we discussed in the introduction, from a local solution of spindle, disk solutions can be obtained by distinct global completion. In our case, it is required to have $A^1=A^2=A^3$, $A^0=0$, and $\lambda_1=\lambda_2=-\lambda_3$ for the disk solutions, \cite{Couzens:2021rlk, Suh:2021hef}. However, then, from \eqref{fourvectors}, the flavor vector fields are trivial and, from \eqref{fixedp}, we do not have the non-trivial vacuum dual to the mass-deformed ABJM theory. 

It would be most interesting to determine the boundary conditions analytically and find analytic expression of the Bekenstein-Hawking entropy. The analytic expressions of the boundary condition would enable us to study the parameter space of the solutions and also to see if the solutions in the twist class are allowed or not.

If the analytic expression of the Bekenstein-Hawking entropy is available, it would be interesting to reproduce it from the field theory calculations as in \cite{Benini:2015noa, Benini:2015eyy} or from the gravitational blocks by generalizing the results of \cite{Faedo:2021nub, Faedo:2022rqx, Boido:2022iye, Boido:2022mbe} beyond the GK geometry.

\bigskip
\leftline{\bf Acknowledgements}
\noindent We thank Krzysztof Pilch for teaching us the fundamentals of gauged $\mathcal{N}=8$ supergravity. We are grateful to I. Arav, J. P. Gauntlett, Y. Jiao, M. M. Roberts, and C. Rosen for pointing out some mistakes in a previous version of the manuscript. Most of this work was carried out while the author was a postdoctoral fellow at Kyung Hee University, Seoul, and Kavli Institute for Theoretical Sciences, Beijing. This work was supported by the National Research Foundation of Korea under the grant NRF-2020R1A2C1008497, Kavli Institute for Theoretical Sciences, Beijing, and Kumoh National Institute of Technology.

\appendix 
\section{The $U(1)^2$-invariant truncation} \label{appA}
\renewcommand{\theequation}{A.\arabic{equation}}
\setcounter{equation}{0} 

In this appendix we review the $U(1)^2$-invariant truncation of $SO(8)$-gauged $\mathcal{N}=8$ supergravity in four dimensions which was studied in \cite{Bobev:2018uxk} and we closely follow \cite{Bobev:2018uxk}. The $U(1)^2$ is the Cartan subgroup of the standard $SU(3)\subset{S}O(6)\subset{S}O(8)$. This truncation is a generalization of the $SU(3)$-invariant truncation, \cite{Warner:1983vz, Warner:1983du, Ahn:2000aq, Ahn:2000mf, Corrado:2001nv, Bobev:2009ms}, and also the $U(1)^4$-invariant truncation, also known as the STU model, \cite{Behrndt:1996hu, Duff:1999gh, Cvetic:1999xp}, where $U(1)^4$ is the Cartan subgroup of $SO(8)$.

The four standard Cartan generators of $SO(8)$ can be denoted by $T_{12}$, $T_{34}$, $T_{56}$ and $T_{78}$ where $T_{ij}$ is the generator of rotation in the $(ij)$-plane with a unit charge. Then the two symmetry generators are
\begin{equation} \label{symgen}
\frac{1}{2}\left(T_{12}-T_{34}\right)\,, \qquad \frac{1}{\sqrt{3}}\left(T_{12}+T_{34}-2T_{56}\right)\,,
\end{equation}
under which the eight gravitini, $\psi_\mu\,^i$, and the supersymmetry parameters, $\epsilon^i$, transform. The two invariant gravitini and the supersymmetry parameters are the chiral $\psi^{7,8}$ and $\epsilon^{7,8}$ and their complex conjugates, $\psi_{7,8}$ and $\epsilon_{7,8}$ of opposite chirality.

From the commutant of the generators in \eqref{symgen}, the unbroken gauge symmetry is the Cartan subgroup, $U(1)^4$, of $SO(8)$. The gauge fields are the graviphoton and three gauge fields from vector multiplet and they are the gauge fields in the STU model. The gauge fields in the symplectic frame, $A^{ij}$, $i,j\,=\,1,\ldots,8$, are specified by the canonical gauge fields, $A^\alpha$, $\alpha\,=\,0,\,\ldots,3$,
\begin{align}
A^{12}\,=\,\frac{1}{2}\left(A^0+A^1-A^2-A^3\right)\,, \qquad A^{34}\,=\,\frac{1}{2}\left(A^0-A^1+A^2-A^3\right)\,, \notag \\
A^{56}\,=\,\frac{1}{2}\left(A^0-A^1-A^2+A^3\right)\,, \qquad A^{78}\,=\,\frac{1}{2}\left(A^0+A^1+A^2+A^3\right)\,.
\end{align}

The scalar 56-bein in the symmetric gauge is given by
\begin{equation}
\mathcal{V}\,\equiv\,\left(
\begin{array}{ll}
 u_{ij}\,^{IJ} & v_{ijKL} \\
 v^{klIJ} & u^{kl}\,_{KL}
\end{array}
\right)\,=\,\text{exp}\left(
\begin{array}{ll}
 0 & \phi_{ijkl} \\
 \phi^{ijkl} & 0
\end{array}
\right)\,\in\,E_{7(7)}/SU(8)\,,
\end{equation}
where we define
\begin{equation}
\phi_{ijkl}\,=\,\frac{1}{24}\epsilon_{ijklmnpq}\phi^{mnpq}\,, \qquad \phi^{ijkl}\,=\,\left(\phi_{ijkl}\right)^*\,,
\end{equation}
which are completely antisymmetric complex self-dual scalar fields. Then the $U(1)^2$-invariant $\phi_{ijkl}$ are given by
\begin{align}
&\phi_{1278}\,=\,-\frac{1}{2}\lambda_1e^{i\varphi_1}\,, \qquad  \phi_{3478}\,=\,-\frac{1}{2}\lambda_2e^{i\varphi_2}\,, \qquad \phi_{5678}\,=\,-\frac{1}{2}\lambda_3e^{i\varphi_3}\,, \notag \\
&\phi_{1234}\,=\,-\frac{1}{2}\lambda_3e^{-i\varphi_3}\,, \qquad  \phi_{1256}\,=\,-\frac{1}{2}\lambda_2e^{-i\varphi_2}\,, \qquad \phi_{3456}\,=\,-\frac{1}{2}\lambda_1e^{-i\varphi_1}\,, \notag \\
&\phi_{1357}\,=\,-\phi_{1467}\,=\,-\phi_{2367}\,=\,-\phi_{2457}\,=\,\frac{1}{4}\left(\chi_1\cos\psi_1+i\chi_2\sin\psi_2\right)\,, \notag \\
&\phi_{1367}\,=\,\phi_{1457}\,=\,\phi_{2357}\,=\,-\phi_{2467}\,=\,-\frac{1}{4}\left(\chi_1\sin\psi_1-i\chi_2\cos\psi_2\right)\,, \notag \\
&\phi_{1368}\,=\,\phi_{1458}\,=\,\phi_{2358}\,=\,-\phi_{2468}\,=\,-\frac{1}{4}\left(\chi_1\cos\psi_1-i\chi_2\sin\psi_2\right)\,, \notag \\
&\phi_{1358}\,=\,-\phi_{1468}\,=\,-\phi_{2368}\,=\,-\phi_{2458}\,=\,-\frac{1}{4}\left(\chi_1\sin\psi_1+i\chi_2\cos\psi_2\right)\,,
\end{align}
where we introduced the scalar fields and they corresponds to the scalar fields, $(\lambda_i,\,\varphi_i)$, from three $\mathcal{N}=2$ vector multiplets and, $(\chi_r,\,\psi_r)$, from the universal hypermultiplet where $i\,=\,1,2,3$ and $r\,=\,1,2$. They parametrize the special K\"ahler and quarternionic K\"ahler manifolds, respectively,
\begin{equation}
\mathcal{M}_V\times\mathcal{M}_H\,=\,\left[\frac{SU(1,1)}{U(1)}\right]^3\times\frac{SU(2,1)}{SU(2)\times{U}(1)}\,.
\end{equation}
In the following, we will restrict ourselves to the case of
\begin{equation}
\chi\,\equiv\,\chi_2\,,\,\,\,\, \psi\,\equiv\,\psi_2\,, \qquad \chi_1\,=\,0\,, \,\,\,\, \psi_1\,=\,\frac{\pi}{2}\,, \qquad \varphi_1\,=\,\varphi_2\,=\,\varphi_3\,=\,\pi\,.
\end{equation}

The bosonic Lagrangian of the truncation is given by
\begin{align} \label{blag}
&e^{-1}\mathcal{L}\,=\,\frac{1}{2}R-\partial_\mu\chi\partial^\mu\chi-\frac{1}{4}\sinh^2\left(2\chi\right)D_\mu\psi{D}^\mu\psi-\sum_{i=1}^3\partial_\mu\lambda_i\partial^\mu\lambda_i-g^2\mathcal{P} \notag \\
&-\frac{1}{4}\left[e^{-2\left(\lambda_1+\lambda_2+\lambda_3\right)}F_{\mu\nu}^0F^{0\mu\nu}+e^{-2\left(\lambda_1-\lambda_2-\lambda_3\right)}F_{\mu\nu}^1F^{1\mu\nu}+e^{2\left(\lambda_1-\lambda_2+\lambda_3\right)}F_{\mu\nu}^2F^{2\mu\nu}+e^{2\left(\lambda_1+\lambda_2-\lambda_3\right)}F_{\mu\nu}^3F^{3\mu\nu}\right]\,,
\end{align}
where we define
\begin{equation}
D\psi\,\equiv\,d\psi-g\left(A^0-A^1-A^2-A^3\right)\,.
\end{equation}
The scalar potential is given by
\begin{equation}
\mathcal{P}\,=\,\frac{1}{2}\left(\frac{\partial{W}}{\partial\chi}\right)^2+\frac{1}{2}\sum_{i=1}^3\left(\frac{\partial{W}}{\partial\lambda_i}\right)^2-\frac{3}{2}W^2\,,
\end{equation}
where the superpotential is
\begin{equation} \label{suppot}
W\,=\,\frac{2e^{2\left(\lambda_1+\lambda_2+\lambda_3\right)}\sinh^2\chi-\cosh^2\chi\left(e^{2\left(\lambda_1+\lambda_2+\lambda_2\right)}+e^{2\lambda_1}+e^{2\lambda_2}+e^{2\lambda_3}\right)}{2e^{\lambda_1+\lambda_2+\lambda_3}}\,.
\end{equation}
The R-symmetry vector field, the massive vector field, and two flavor symmetry vector fields are given by, respectively,
\begin{align}
A_R\,=&\,-g\left(A^0+A^1+A^2+A^3\right)\,, \notag \\
A_m\,=&\,g\left(A^0-A^1-A^2-A^3\right)\,, \notag \\
A_{F_1}\,=&\,\frac{1}{2}g\left(A^1-A^2\right)\,, \notag \\
A_{F_2}\,=&\,\frac{1}{\sqrt{3}}g\left(A^1+A^2-2A^3\right)\,,
\end{align}
where the normalizations of the flavor symmetries are determined by \eqref{symgen}.

The supersymmetry variations of fermionic fields are given by
\begin{align}
\delta\chi^{ijk}\,=&\,-\mathcal{A}_\mu\,^{ijkl}\gamma^{\mu}\epsilon_l+\frac{3}{2}\gamma^{\mu\nu}\overline{F}_{\mu\nu}^{-[ij}\epsilon^{k]}-2gA_{2l}\,^{ijk}\epsilon^l\,,\notag \\
\delta\psi_\mu\,^i\,=&\,2D_\mu\epsilon^i+\frac{\sqrt{2}}{4}\overline{F}_{\nu\rho}^{-ij}\gamma^{\nu\rho}\gamma_\mu\epsilon_j+\sqrt{2}gA_1\,^{ij}\gamma_\mu\epsilon_j\,.
\end{align}
Refer \cite{deWit:1986oxb} for the definitions of the tensors, $\mathcal{A}_\mu\,^{ijkl}$, $A_1\,^{ij}$ and $A_{2l}\,^{ijk}$. The scalar-field dressed field strengths, $\overline{F}^{-ij}$, are defined by
\begin{equation}
F^{-IJ}\,=\,\left(u_{ij}\,^{IJ}+v_{ijIJ}\right)\overline{F}^{-ij}\,, \qquad F^{-IJ}\,=\,\frac{1}{2}\left(F^{IJ}-i*F^{IJ}\right)\,,
\end{equation}
and the minus sign denotes the anti-self dual part of the field strength. 

In the $U(1)^2$-invariant truncation, we find the supersymmetry parameters to be
\begin{equation}
\epsilon^1\,=\,\ldots\,=\,\epsilon^6\,=\,0\,,
\end{equation} 
and the only non-trivial parameters are $\epsilon^7$ and $\epsilon^8$ with $\epsilon_i\,=\,\left(\epsilon^i\right)^*$. We find the real superpotential, \eqref{suppot}, from the eigenvalues of $A_1\,^{ij}$ tensors,
\begin{equation}
A_1\,^{77}\epsilon_7\,=\,-\frac{1}{2}W\epsilon_7\,, \qquad A_1\,^{88}\epsilon_8\,=\,-\frac{1}{2}W\epsilon_8\,.
\end{equation}
We also define $B_\mu$ and $H_{\mu\nu}$ from
\begin{align}
B_\mu\,\equiv&\,B_\mu\,^7\,_8\,=\,-B_\mu\,^8\,_7, \notag \\
H_{\mu\nu}\,\equiv&\,\overline{F}_{\mu\nu}^{78}\,,
\end{align}
where explicitly we have
\begin{align} \label{defhb}
H_{\mu\nu}\,\equiv&\,\overline{F}_{\mu\nu}^{78}\,=\,\frac{1}{2}\left(e^{-\lambda_1-\lambda_2-\lambda_3}F_{\mu\nu}^{0}+e^{-\lambda_1+\lambda_2+\lambda_3}F_{\mu\nu}^{1}+e^{\lambda_1-\lambda_2+\lambda_3}F_{\mu\nu}^{2}+e^{\lambda_1+\lambda_2-\lambda_3}F_{\mu\nu}^{3}\right)\,, \notag \\
B_\mu\,\equiv&\,-g\left(A^0_\mu+A^1_\mu+A^2_\mu+A^3_\mu\right)-\frac{1}{2}\left(\cosh\left(2\chi\right)-1\right)D_\mu\psi\,.
\end{align}

Instead of the Weyl spinors, $\epsilon^i$, we rewrite the supersymmetry variations with a complex Dirac spinor,
\begin{equation}
\epsilon\,=\,\left(\epsilon_7+\epsilon^7\right)+i\left(\epsilon_8+\epsilon^8\right)\,.
\end{equation}
Similar change was made in, $e.g.$, \cite{Anabalon:2022aig}. The spin-3/2 field variation reduces to
\begin{equation} \label{spin3/2v}
\left[2\nabla_\mu-iB_\mu-\frac{g}{\sqrt{2}}W\gamma_\mu-i\frac{1}{2\sqrt{2}}H_{\nu\rho}\gamma^{\nu\rho}\gamma_\mu\right]\epsilon\,=\,0\,.
\end{equation}
The spin-1/2 field variation reduces to the gaugino variations,
\begin{align} \label{gauginov}
\left[\gamma^\mu\partial_\mu\lambda_1+\frac{g}{\sqrt{2}}\partial_{\lambda_1}W+i\frac{1}{2\sqrt{2}}\gamma^{\mu\nu}\overline{F}_{\mu\nu}^{12}\right]\epsilon\,=\,0\,, \notag \\
\left[\gamma^\mu\partial_\mu\lambda_2+\frac{g}{\sqrt{2}}\partial_{\lambda_2}W+i\frac{1}{2\sqrt{2}}\gamma^{\mu\nu}\overline{F}_{\mu\nu}^{34}\right]\epsilon\,=\,0\,, \notag \\
\left[\gamma^\mu\partial_\mu\lambda_3+\frac{g}{\sqrt{2}}\partial_{\lambda_3}W+i\frac{1}{2\sqrt{2}}\gamma^{\mu\nu}\overline{F}_{\mu\nu}^{56}\right]\epsilon\,=\,0\,,
\end{align}
and the hyperino variations,
\begin{equation} \label{hyperinov}
\left[\gamma^\mu\partial_\mu\chi+\frac{g}{\sqrt{2}}\partial_\chi{W}+i\frac{1}{2}\partial_\chi{B}_\mu\gamma^\mu\right]\epsilon\,=\,0\,,
\end{equation}
where we have
\begin{align}
F^{0}\,=&\,\frac{1}{2}e^{\lambda_1+\lambda_2+\lambda_3}\left(\overline{F}^{12}+\overline{F}^{34}+\overline{F}^{56}+\overline{F}^{78}\right)\,, \notag \\
F^{1}\,=&\,\frac{1}{2}e^{\lambda_1-\lambda_2-\lambda_3}\left(\overline{F}^{12}-\overline{F}^{34}-\overline{F}^{56}+\overline{F}^{78}\right)\,, \notag \\
F^{2}\,=&\,-\frac{1}{2}e^{-\lambda_1+\lambda_2-\lambda_3}\left(\overline{F}^{12}-\overline{F}^{34}+\overline{F}^{56}-\overline{F}^{78}\right)\,, \notag \\
F^{3}\,=&\,-\frac{1}{2}e^{-\lambda_1-\lambda_2+\lambda_3}\left(\overline{F}^{12}+\overline{F}^{34}-\overline{F}^{56}-\overline{F}^{78}\right)\,.
\end{align}

We present the equations of motion from the Lagrangian in \eqref{blag}. The Einstein equations are
\begin{align}
R_{\mu\nu}-&\frac{1}{2}Rg_{\mu\nu}+g^2\mathcal{P}g_{\mu\nu}-2\left(T_{\mu\nu}^\chi+T_{\mu\nu}^{\lambda_1}+T_{\mu\nu}^{\lambda_2}+T_{\mu\nu}^{\lambda_3}\right)-\frac{1}{2}\sinh^2\left(2\chi\right)T_{\mu\nu}^\psi \notag \\
-&e^{-2\left(\lambda_1+\lambda_2+\lambda_3\right)}T_{\mu\nu}^{A^0}-e^{-2\left(\lambda_1-\lambda_2-\lambda_3\right)}T_{\mu\nu}^{A^1}-e^{2\left(\lambda_1-\lambda_2+\lambda_3\right)}T_{\mu\nu}^{A^2}-e^{2\left(\lambda_1+\lambda_2-\lambda_3\right)}T_{\mu\nu}^{A^3}\,=\,0\,,
\end{align}
with the energy-momentum tensors by
\begin{align}
T_{\mu\nu}^X\,=&\,\partial_\mu{X}\partial_\nu{X}-\frac{1}{2}g_{\mu\nu}\partial_\rho{X}\partial^\rho{X}\,, \notag \\
T_{\mu\nu}^{A^\alpha}\,=&\,g^{\rho\sigma}F_{\mu\rho}^\alpha{F}_{\nu\sigma}^\alpha-\frac{1}{4}g_{\mu\nu}F_{\rho\sigma}^\alpha{F}^{\alpha\rho\sigma}\,,
\end{align}
where $X$ denotes a scalar field. The Maxwell equations are
\begin{align}
\partial_\nu\left(\sqrt{-g}e^{-2\left(\lambda_1+\lambda_2+\lambda_3\right)}F^{0\mu\nu}\right)-\frac{1}{2}\sqrt{-g}\sinh^2\left(2\chi\right)g^{\mu\nu}D_\nu\psi\,=&\,0\,, \notag \\
\partial_\nu\left(\sqrt{-g}e^{-2\left(\lambda_1-\lambda_2-\lambda_3\right)}F^{1\mu\nu}\right)+\frac{1}{2}\sqrt{-g}\sinh^2\left(2\chi\right)g^{\mu\nu}D_\nu\psi\,=&\,0\,, \notag \\
\partial_\nu\left(\sqrt{-g}e^{2\left(\lambda_1-\lambda_2+\lambda_3\right)}F^{2\mu\nu}\right)+\frac{1}{2}\sqrt{-g}\sinh^2\left(2\chi\right)g^{\mu\nu}D_\nu\psi\,=&\,0\,, \notag \\
\partial_\nu\left(\sqrt{-g}e^{2\left(\lambda_1+\lambda_2-\lambda_3\right)}F^{3\mu\nu}\right)+\frac{1}{2}\sqrt{-g}\sinh^2\left(2\chi\right)g^{\mu\nu}D_\nu\psi\,=&\,0\,,
\end{align}
and the scalar field equations are given by
\begin{align}
&\frac{1}{\sqrt{-g}}\partial_\mu\left(\sqrt{-g}g^{\mu\nu}\partial_\nu\lambda_1\right)-\frac{g^2}{2}\frac{\partial\mathcal{P}}{\partial\lambda_1} \notag \\
&+\frac{1}{4}e^{-2\left(\lambda_1+\lambda_2+\lambda_3\right)}F_{\mu\nu}^0F^{0\mu\nu}+\frac{1}{4}e^{-2\left(\lambda_1-\lambda_2-\lambda_3\right)}F_{\mu\nu}^1F^{1\mu\nu} \notag \\
&-\frac{1}{4}e^{2\left(\lambda_1-\lambda_2+\lambda_3\right)}F_{\mu\nu}^2F^{2\mu\nu}-\frac{1}{4}e^{2\left(\lambda_1+\lambda_2-\lambda_3\right)}F_{\mu\nu}^3F^{3\mu\nu}\,=\,0\,, \notag \\  \notag \\
&\frac{1}{\sqrt{-g}}\partial_\mu\left(\sqrt{-g}g^{\mu\nu}\partial_\nu\lambda_2\right)-\frac{g^2}{2}\frac{\partial\mathcal{P}}{\partial\lambda_2} \notag \\
&+\frac{1}{4}e^{-2\left(\lambda_1+\lambda_2+\lambda_3\right)}F_{\mu\nu}^0F^{0\mu\nu}-\frac{1}{4}e^{-2\left(\lambda_1-\lambda_2-\lambda_3\right)}F_{\mu\nu}^1F^{1\mu\nu} \notag \\
&+\frac{1}{4}e^{2\left(\lambda_1-\lambda_2+\lambda_3\right)}F_{\mu\nu}^2F^{2\mu\nu}-\frac{1}{4}e^{2\left(\lambda_1+\lambda_2-\lambda_3\right)}F_{\mu\nu}^3F^{3\mu\nu}\,=\,0\,, \notag \\  \notag \\
&\frac{1}{\sqrt{-g}}\partial_\mu\left(\sqrt{-g}g^{\mu\nu}\partial_\nu\lambda_3\right)-\frac{g^2}{2}\frac{\partial\mathcal{P}}{\partial\lambda_3} \notag \\
&+\frac{1}{4}e^{-2\left(\lambda_1+\lambda_2+\lambda_3\right)}F_{\mu\nu}^0F^{0\mu\nu}-\frac{1}{4}e^{-2\left(\lambda_1-\lambda_2-\lambda_3\right)}F_{\mu\nu}^1F^{1\mu\nu} \notag \\
&-\frac{1}{4}e^{2\left(\lambda_1-\lambda_2+\lambda_3\right)}F_{\mu\nu}^2F^{2\mu\nu}+\frac{1}{4}e^{2\left(\lambda_1+\lambda_2-\lambda_3\right)}F_{\mu\nu}^3F^{3\mu\nu}\,=\,0\,,
\end{align}
with
\begin{equation}
\frac{1}{\sqrt{-g}}\partial_\mu\left(\sqrt{-g}g^{\mu\nu}\partial_\nu\chi\right)-\frac{g^2}{2}\frac{\partial\mathcal{P}}{\partial\chi}+\frac{1}{8}\sinh\left(4\chi\right)D_\mu\psi{D}^\mu\psi\,=\,0\,.
\end{equation}

\vspace{1.6cm}

\subsection{Truncation to the STU model}

The truncation reduces to the STU model, \cite{Behrndt:1996hu, Duff:1999gh, Cvetic:1999xp}, by setting the charged complex scalar field to vanish,
\begin{equation}
\chi\,=\,0\,, \qquad \psi\,=\,\frac{\pi}{2}\,,
\end{equation}
and the scalar potential is
\begin{equation}
\mathcal{P}\,=\,-2\left[\cosh\left(2\lambda_1\right)+\cosh\left(2\lambda_2\right)+\cosh\left(2\lambda_3\right)\right]\,.
\end{equation}
Reparametrizing $\lambda_i\rightarrow\frac{1}{2}\varphi_I$ and $F^\alpha\rightarrow\frac{1}{\sqrt{2}}F^I$ with $g=\frac{1}{\sqrt{2}}$, we obtain the normalizations of the STU model used in \cite{Ferrero:2021etw}.

When we further impose
\begin{equation}
\lambda_i\,=\,0\,,\qquad \varphi_i\,=\,\pi\,, \qquad A\,\equiv\,A^0\,=\,A^1\,=\,A^2\,=\,A^3\,,
\end{equation}
we obtain the Lagrangian of minimal gauged supergravity,
\begin{equation}
e^{-1}\mathcal{L}\,=\,\frac{1}{2}R-6g^2-F_{\mu\nu}F^{\mu\nu}\,,
\end{equation}
where $F=dA$. We set $g=\frac{1}{\sqrt{2}}$ and $F\,\rightarrow\frac{1}{\sqrt{2}}F$ and obtain the Lagrangian of the normalization of minimal gauged supergravity employed in \cite{Ferrero:2020twa}.

\subsection{Truncation to minimal gauged supergravity via W}

There is an alternative truncation to minimal gauged supergravity associated with the Warner $\mathcal{N}=2$ vacuum. We have the scalar fields to be at their values of the Warner $\mathcal{N}=2$ vacuum and impose the gauge fields to be
\begin{align}
&\tanh\chi\,=\,\frac{1}{\sqrt{3}}\,, \qquad \psi\,=\,\frac{\pi}{2}\,, \qquad \tanh\lambda_i\,=\,2-\sqrt{3}\,, \qquad \varphi_i\,=\,\pi\,, \notag \\
&\frac{1}{3}A\,\equiv\,\frac{1}{3}A^0\,=\,A^1\,=\,A^2\,=\,A^3\,,
\end{align}
and find
\begin{equation}
e^{-1}\mathcal{L}\,=\,\frac{1}{2}R-\frac{9\sqrt{3}}{2}g^2-\frac{1}{12\sqrt{3}}F_{\mu\nu}F^{\mu\nu}\,,
\end{equation}
where $F=dA$. When we have $g=\frac{\sqrt{2}}{3^{3/4}}$ and $F\rightarrow\sqrt{2}\,3^{3/4}F$, it reduces to the action of minimal gauged supergravity employed in \cite{Ferrero:2020twa}.

\section{Supersymmetry variations} \label{appB}
\renewcommand{\theequation}{B.\arabic{equation}}
\setcounter{equation}{0} 

\subsection{Derivation of the BPS equations}

Following the arguments in \cite{Arav:2022lzo}, we derive the BPS equations required to construct the spindle solutions. We consider the background with the gauge fields,
\begin{align}
ds^2\,=&\,e^{2V}ds_{AdS_2}^2+f^2dy^2+h^2dz^2\,, \notag \\
A^\alpha\,=&\,a^\alpha{d}z\,,
\end{align}
where $V$, $f$, $h$, and $a^\alpha$, $\alpha\,=\,0,\,\ldots,\,3$, are functions of coordinate $y$ only. We employ the gamma matrices,
\begin{equation}
\gamma^m\,=\,\Gamma^m\otimes\sigma^3\,, \qquad \gamma^2\,=\,\mathbb{I}_2\otimes\sigma^1\,, \qquad \gamma^3\,=\,\mathbb{I}_2\otimes\sigma^2\,,
\end{equation}
and the spinors,
\begin{equation}
\epsilon\,=\,\psi\otimes\chi\,,
\end{equation}
where $\Gamma^m$ are two-dimensional gamma matrices of mostly plus signature. The two-dimensional spinor satisfies
\begin{equation}
D_m\psi\,=\,\frac{1}{2}\kappa\Gamma_m\psi\,,
\end{equation}
where $\kappa\,=\,\pm1$.

From the directions tangent to $AdS_2$, the gravitino variation reduces to
\begin{equation}
\left[-i\left(\sqrt{2}\kappa{e}^{-V}+H_{23}\right)\gamma^{23}+\sqrt{2}V'f^{-1}\gamma^2\right]\epsilon\,=\,gW\epsilon\,.
\end{equation}
It requires a projection condition,
\begin{equation}
\left[i\cos\xi\gamma^{23}+\sin\xi\gamma^2\right]\epsilon\,=\,+\epsilon\,,
\end{equation}
where we introduce $\xi$ from
\begin{equation} \label{bfive}
-\sqrt{2}\kappa{e}^{-V}-H_{23}\,=\,gW\cos\xi\,, \qquad \sqrt{2}V'f^{-1}\,=\,gW\sin\xi\,.
\end{equation}
The projection condition is solved by
\begin{equation}
\epsilon\,=\,e^{i\frac{\xi}{2}\gamma^3}\eta\,, \qquad \gamma^2\eta\,=\,+i\gamma^3\eta\,.
\end{equation}
We find $\partial_z\xi\,=\,0$ from \eqref{bfive}. At $\xi\,=\,0,\,\pi$, the spinors have definite chirality with respect to $\gamma^{23}$,
\begin{equation}
\xi\,=\,0,\pi\,, \qquad \gamma^{23}\epsilon\,=\,\pm{i}\epsilon\,.
\end{equation}

From the $y$ direction, the spin-3/2 field variation, \eqref{spin3/2v}, reduces to
\begin{equation} \label{32y}
\left[\partial_y-\frac{1}{2}V'+\frac{i}{2}\left(\partial_y\xi+\sqrt{2}fH_{23}+\kappa{f}e^{-V}\right)\gamma^3\right]\eta\,=\,0\,,
\end{equation}
where we employed \eqref{bfive}. From the $z$ direction, we find
\begin{align} \label{32z}
\Big[2\partial_z-iB_z&+if^{-1}h'\cos\xi-i\frac{1}{\sqrt{2}}H_{23}h\sin\xi \notag \\
&+\left(f^{-1}h'\sin\xi-\frac{1}{\sqrt{2}}gWh+\frac{1}{\sqrt{2}}H_{23}h\cos\xi\right)\gamma^3\Big]\eta\,=\,0\,.
\end{align}

When we have $\left(a_1+ia_2\gamma^3\right)\eta\,=\,0$, it requires $a_1^2+a_2^2\,=\,0$. Hence, we find from \eqref{32y} and \eqref{32z},
\begin{equation}
\eta\,=\,e^{V/2}e^{isz}\eta_0\,,
\end{equation}
where $\eta_0$ is independent of $y$ and $z$ and we obtain
\begin{align}
\partial_y\xi+\sqrt{2}fH_{23}+\kappa{f}e^{-V}\,&=\,0\,, \notag \\
\left(s-B_z\right)+f^{-1}h'\cos\xi-\frac{1}{\sqrt{2}}H_{23}h\sin\xi\,&=\,0\,, \notag \\
f^{-1}h'\sin\xi-\frac{gWh}{\sqrt{2}}+\frac{1}{\sqrt{2}}H_{23}h\cos\xi\,&=\,0\,.
\end{align}
Thus we find
\begin{align}
f^{-1}h'\,=&\,\frac{gWh}{\sqrt{2}}\sin\xi-\left(s-B_z\right)\cos\xi\,, \notag \\
hH_{23}\,=&\,gWh\cos\xi+\sqrt{2}\left(s-B_z\right)\sin\xi\,,
\end{align}
and, from the first constraint in \eqref{bfive}, we obtain
\begin{equation}
\left(s-B_z\right)\sin\xi\,=\,-\sqrt{2}gWh\cos\xi-\kappa{h}e^{-V}\,,
\end{equation}
and thus
\begin{align}
H_{23}\,=&\,-gW\cos\xi-\sqrt{2}\kappa{e}^{-V}\,, \notag \\
f^{-1}\partial_y\xi\,=&\,\sqrt{2}gW\cos\xi+\kappa{e}^{-V}\,.
\end{align}
If we have $\xi\ne0$, by solving for $\left(s-B_z\right)$, we also obtain 
\begin{equation}
f^{-1}\frac{h'}{h}\sin\xi\,=\,\kappa{e}^{-V}\cos\xi+\frac{gW}{\sqrt{2}}\left(1+\cos^2\xi\right)\,.
\end{equation}

From the gaugino variations, \eqref{gauginov}, in a similar way, we obtain
\begin{equation}
f^{-1}\lambda_i'+\frac{g}{\sqrt{2}}\partial_{\lambda_i}W\sin\xi\,=\,0\,,
\end{equation}
with
\begin{align}
g\partial_{\lambda_1}W\cos\xi+\overline{F}_{23}^{12}\,=\,0\,, \notag \\
g\partial_{\lambda_2}W\cos\xi+\overline{F}_{23}^{34}\,=\,0\,, \notag \\
g\partial_{\lambda_3}W\cos\xi+\overline{F}_{23}^{56}\,=\,0\,.
\end{align}

From the hyperino variations, \eqref{hyperinov}, we obtain
\begin{align}
f^{-1}\chi'\sin\xi+\frac{g}{\sqrt{2}}\partial_\chi{W}\,=&\,0\,, \notag \\
\sqrt{2}g\partial_\chi{W}\cos\xi-\partial_\chi{B}_z\sin\xi{h}^{-1}\,=&\,0\,.
\end{align}

{\bf Summary:} If we have $\sin\xi\ne0$, the complete BPS equations are given by
\begin{align}
f^{-1}\xi'\,=&\,\sqrt{2}gW\cos\xi+\kappa{e}^{-V}\,, \notag \\
f^{-1}V'\,=&\,\frac{g}{\sqrt{2}}W\sin\xi\,, \notag \\
f^{-1}\lambda_i'\,=&\,-\frac{g}{\sqrt{2}}\partial_{\lambda_i}W\sin\xi\,, \notag \\
f^{-1}\chi'\,=&\,-\frac{g}{\sqrt{2}}\frac{\partial_\chi{W}}{\sin\xi}\,, \notag \\
f^{-1}\frac{h'}{h}\sin\xi\,=&\,\kappa{e}^{-V}\cos\xi+\frac{gW}{\sqrt{2}}\left(1+\cos^2\xi\right)\,, 
\end{align}
with two constraints,
\begin{align}
\left(s-B_z\right)\sin\xi\,=&\,-\sqrt{2}gWh\cos\xi-\kappa{h}e^{-V}\,, \notag \\
\sqrt{2}g\partial_\chi{W}\cos\xi\,=&\,\partial_\chi{B}_z\sin\xi{h}^{-1}\,.
\end{align}
The scalar-field dressed field strengths are given by
\begin{align}
\overline{F}_{23}^{12}\,=&\,-g\partial_{\lambda_1}W\cos\xi\,, \notag \\
\overline{F}_{23}^{34}\,=&\,-g\partial_{\lambda_2}W\cos\xi\,, \notag \\
\overline{F}_{23}^{56}\,=&\,-g\partial_{\lambda_3}W\cos\xi\,, \notag \\
H_{23}\,=&\,-gW\cos\xi-\sqrt{2}\kappa{e}^{-V}\,.
\end{align}
We have checked that the BPS equations solve the equations of motion from the Lagrangian in \eqref{blag} as presented in appendix \ref{appA}.

We also obtain
\begin{equation}
\partial_yW\,=\,-\frac{g}{\sqrt{2}}f\sin\xi\left[\sum_{i=1}^3\left(\partial_{\lambda_i}W\right)^2+\frac{1}{\sin^2\xi}\left(\partial_\chi{W}\right)^2\right]\,,
\end{equation}
and, thus, the superpotential, $W$, is monotonic along the BPS flow if the sign of $f\sin\xi$ does not change.

We find an integral of the BPS equations,
\begin{equation}
he^{-V}\,=\,k\sin\xi\,,
\end{equation}
where $k$ is a constant. Employing this to eliminate $h$, we find the BPS equations to be
\begin{align}
f^{-1}\xi'\,=&\,-k^{-1}\left(s-B_z\right)e^{-V}\,, \notag \\
f^{-1}V'\,=&\,\frac{g}{\sqrt{2}}W\sin\xi\,, \notag \\
f^{-1}\lambda_i'\,=&\,-\frac{g}{\sqrt{2}}\partial_{\lambda_i}W\sin\xi\,, \notag \\
f^{-1}\chi'\,=&\,-\frac{g}{\sqrt{2}}\frac{\partial_\chi{W}}{\sin\xi}\,,
\end{align}
with the two constraints,
\begin{align} \label{twoconstraints}
\left(s-B_z\right)\,=&\,-k\left(\sqrt{2}gWe^V\cos\xi+\kappa\right)\,, \notag \\
\sqrt{2}g\partial_\chi{W}\cos\xi\,=&\,k^{-1}e^{-V}\partial_\chi{B}_z\,.
\end{align}

From the definition of $B_z$ in \eqref{defhb}, we find
\begin{equation}
\partial_\chi{B}_z\,=\,-\sinh\left(2\chi\right)D_z\psi\,.
\end{equation}
If we have $\chi\ne0$, from the second constraint in \eqref{twoconstraints}, we find
\begin{equation} \label{dspsidzpsi}
D_z\psi\,=\,-\frac{\sqrt{2}gke^V\partial_\chi{W}\cos\xi}{\sinh\left(2\chi\right)}\,,
\end{equation}
and the right hand side is independent of $\chi$. By differentiating \eqref{dspsidzpsi} we find expressions of fluxes in terms of the derivatives of other fields,
\begin{equation}
F^\alpha_{yz}\,=\,\left(a^\alpha\right)'\,=\,\left(\mathcal{I}^{(\alpha)}\right)'\,,
\end{equation}
where we have
\begin{align}
\mathcal{I}^{(0)}\,\equiv&\,\frac{1}{\sqrt{2}}ke^V\cos\xi\,e^{\lambda_1+\lambda_2+\lambda_3}\,, \notag \\
\mathcal{I}^{(1)}\,\equiv&\,\frac{1}{\sqrt{2}}ke^V\cos\xi\,e^{\lambda_1-\lambda_2-\lambda_3}\,, \notag \\
\mathcal{I}^{(2)}\,\equiv&\,\frac{1}{\sqrt{2}}ke^V\cos\xi\,e^{-\lambda_1+\lambda_2-\lambda_3}\,, \notag \\
\mathcal{I}^{(3)}\,\equiv&\,\frac{1}{\sqrt{2}}ke^V\cos\xi\,e^{-\lambda_1-\lambda_2+\lambda_3}\,.
\end{align}

There is a symmetry of the BPS equations,
\begin{equation} \label{hsymm}
h\,\rightarrow\,-h\,, \qquad z\,\rightarrow\,-z\,,
\end{equation}
when we have $B_z\rightarrow-B_z$, $s\rightarrow-s$, $a^\alpha\rightarrow-a^\alpha$, $k\rightarrow-k$ and $F_{23}^\alpha\rightarrow+F_{23}^\alpha$. The frame is invariant under this transformation. We fix $h\ge0$ by this symmetry in the main text.

\bibliographystyle{JHEP}
\bibliography{reference_20221008}

\providecommand{\href}[2]{#2}\begingroup\raggedright\begin{thebibliography}{10}

\bibitem{Witten:1988ze}
E.~Witten, \emph{{Topological Quantum Field Theory}},
  \href{http://dx.doi.org/10.1007/BF01223371}{\emph{Commun. Math. Phys.} {\bf
  117} (1988) 353}.

\bibitem{Maldacena:2000mw}
J.~M. Maldacena and C.~Nunez, \emph{{Supergravity description of field theories
  on curved manifolds and a no go theorem}},
  \href{http://dx.doi.org/10.1142/S0217751X01003937}{\emph{Int. J. Mod. Phys.
  A} {\bf 16} (2001) 822--855},
  [\href{https://arxiv.org/abs/hep-th/0007018}{{\tt hep-th/0007018}}].

\bibitem{Ferrero:2020laf}
P.~Ferrero, J.~P. Gauntlett, J.~M. P\'erez Ipi\~na, D.~Martelli and J.~Sparks,
  \emph{{D3-Branes Wrapped on a Spindle}},
  \href{http://dx.doi.org/10.1103/PhysRevLett.126.111601}{\emph{Phys. Rev.
  Lett.} {\bf 126} (2021) 111601},
  [\href{https://arxiv.org/abs/2011.10579}{{\tt 2011.10579}}].

\bibitem{Hosseini:2021fge}
S.~M. Hosseini, K.~Hristov and A.~Zaffaroni, \emph{{Rotating multi-charge
  spindles and their microstates}},
  \href{http://dx.doi.org/10.1007/JHEP07(2021)182}{\emph{JHEP} {\bf 07} (2021)
  182}, [\href{https://arxiv.org/abs/2104.11249}{{\tt 2104.11249}}].

\bibitem{Boido:2021szx}
A.~Boido, J.~M.~P. Ipi\~na and J.~Sparks, \emph{{Twisted D3-brane and M5-brane
  compactifications from multi-charge spindles}},
  \href{http://dx.doi.org/10.1007/JHEP07(2021)222}{\emph{JHEP} {\bf 07} (2021)
  222}, [\href{https://arxiv.org/abs/2104.13287}{{\tt 2104.13287}}].

\bibitem{Ferrero:2020twa}
P.~Ferrero, J.~P. Gauntlett, J.~M.~P. Ipi\~na, D.~Martelli and J.~Sparks,
  \emph{{Accelerating black holes and spinning spindles}},
  \href{http://dx.doi.org/10.1103/PhysRevD.104.046007}{\emph{Phys. Rev. D} {\bf
  104} (2021) 046007}, [\href{https://arxiv.org/abs/2012.08530}{{\tt
  2012.08530}}].

\bibitem{Cassani:2021dwa}
D.~Cassani, J.~P. Gauntlett, D.~Martelli and J.~Sparks, \emph{{Thermodynamics
  of accelerating and supersymmetric AdS4 black holes}},
  \href{http://dx.doi.org/10.1103/PhysRevD.104.086005}{\emph{Phys. Rev. D} {\bf
  104} (2021) 086005}, [\href{https://arxiv.org/abs/2106.05571}{{\tt
  2106.05571}}].

\bibitem{Ferrero:2021ovq}
P.~Ferrero, M.~Inglese, D.~Martelli and J.~Sparks, \emph{{Multicharge
  accelerating black holes and spinning spindles}},
  \href{http://dx.doi.org/10.1103/PhysRevD.105.126001}{\emph{Phys. Rev. D} {\bf
  105} (2022) 126001}, [\href{https://arxiv.org/abs/2109.14625}{{\tt
  2109.14625}}].

\bibitem{Couzens:2021rlk}
C.~Couzens, K.~Stemerdink and D.~van~de Heisteeg, \emph{{M2-branes on discs and
  multi-charged spindles}},
  \href{http://dx.doi.org/10.1007/JHEP04(2022)107}{\emph{JHEP} {\bf 04} (2022)
  107}, [\href{https://arxiv.org/abs/2110.00571}{{\tt 2110.00571}}].

\bibitem{Ferrero:2021wvk}
P.~Ferrero, J.~P. Gauntlett, D.~Martelli and J.~Sparks, \emph{{M5-branes
  wrapped on a spindle}},
  \href{http://dx.doi.org/10.1007/JHEP11(2021)002}{\emph{JHEP} {\bf 11} (2021)
  002}, [\href{https://arxiv.org/abs/2105.13344}{{\tt 2105.13344}}].

\bibitem{Faedo:2021nub}
F.~Faedo and D.~Martelli, \emph{{D4-branes wrapped on a spindle}},
  \href{http://dx.doi.org/10.1007/JHEP02(2022)101}{\emph{JHEP} {\bf 02} (2022)
  101}, [\href{https://arxiv.org/abs/2111.13660}{{\tt 2111.13660}}].

\bibitem{Giri:2021xta}
S.~Giri, \emph{{Black holes with spindles at the horizon}},
  \href{http://dx.doi.org/10.1007/JHEP06(2022)145}{\emph{JHEP} {\bf 06} (2022)
  145}, [\href{https://arxiv.org/abs/2112.04431}{{\tt 2112.04431}}].

\bibitem{Arav:2022lzo}
I.~Arav, J.~P. Gauntlett, M.~M. Roberts and C.~Rosen, \emph{{Leigh-Strassler
  compactified on a spindle}},
  \href{http://dx.doi.org/10.1007/JHEP10(2022)067}{\emph{JHEP} {\bf 10} (2022)
  067}, [\href{https://arxiv.org/abs/2207.06427}{{\tt 2207.06427}}].

\bibitem{Couzens:2022yiv}
C.~Couzens and K.~Stemerdink, \emph{{Universal spindles: D2's on $\Sigma$ and
  M5's on $\Sigma\times \mathbb{H}^3$}},
  \href{https://arxiv.org/abs/2207.06449}{{\tt 2207.06449}}.

\bibitem{Ferrero:2021etw}
P.~Ferrero, J.~P. Gauntlett and J.~Sparks, \emph{{Supersymmetric spindles}},
  \href{http://dx.doi.org/10.1007/JHEP01(2022)102}{\emph{JHEP} {\bf 01} (2022)
  102}, [\href{https://arxiv.org/abs/2112.01543}{{\tt 2112.01543}}].

\bibitem{Couzens:2021cpk}
C.~Couzens, \emph{{A tale of (M)2 twists}},
  \href{http://dx.doi.org/10.1007/JHEP03(2022)078}{\emph{JHEP} {\bf 03} (2022)
  078}, [\href{https://arxiv.org/abs/2112.04462}{{\tt 2112.04462}}].

\bibitem{Bah:2021mzw}
I.~Bah, F.~Bonetti, R.~Minasian and E.~Nardoni, \emph{{Holographic Duals of
  Argyres-Douglas Theories}},
  \href{http://dx.doi.org/10.1103/PhysRevLett.127.211601}{\emph{Phys. Rev.
  Lett.} {\bf 127} (2021) 211601},
  [\href{https://arxiv.org/abs/2105.11567}{{\tt 2105.11567}}].

\bibitem{Bah:2021hei}
I.~Bah, F.~Bonetti, R.~Minasian and E.~Nardoni, \emph{{M5-brane sources,
  holography, and Argyres-Douglas theories}},
  \href{http://dx.doi.org/10.1007/JHEP11(2021)140}{\emph{JHEP} {\bf 11} (2021)
  140}, [\href{https://arxiv.org/abs/2106.01322}{{\tt 2106.01322}}].

\bibitem{Argyres:1995jj}
P.~C. Argyres and M.~R. Douglas, \emph{{New phenomena in SU(3) supersymmetric
  gauge theory}},
  \href{http://dx.doi.org/10.1016/0550-3213(95)00281-V}{\emph{Nucl. Phys. B}
  {\bf 448} (1995) 93--126}, [\href{https://arxiv.org/abs/hep-th/9505062}{{\tt
  hep-th/9505062}}].

\bibitem{Couzens:2022yjl}
C.~Couzens, H.~Kim, N.~Kim and Y.~Lee, \emph{{Holographic duals of M5-branes on
  an irregularly punctured sphere}},
  \href{http://dx.doi.org/10.1007/JHEP07(2022)102}{\emph{JHEP} {\bf 07} (2022)
  102}, [\href{https://arxiv.org/abs/2204.13537}{{\tt 2204.13537}}].

\bibitem{Bah:2022yjf}
I.~Bah, F.~Bonetti, E.~Nardoni and T.~Waddleton, \emph{{Aspects of irregular
  punctures via holography}},
  \href{http://dx.doi.org/10.1007/JHEP11(2022)131}{\emph{JHEP} {\bf 11} (2022)
  131}, [\href{https://arxiv.org/abs/2207.10094}{{\tt 2207.10094}}].

\bibitem{Couzens:2021tnv}
C.~Couzens, N.~T. Macpherson and A.~Passias, \emph{{$ \mathcal{N} $ = (2, 2)
  AdS$_{3}$ from D3-branes wrapped on Riemann surfaces}},
  \href{http://dx.doi.org/10.1007/JHEP02(2022)189}{\emph{JHEP} {\bf 02} (2022)
  189}, [\href{https://arxiv.org/abs/2107.13562}{{\tt 2107.13562}}].

\bibitem{Suh:2021ifj}
M.~Suh, \emph{{D3-branes and M5-branes wrapped on a topological disc}},
  \href{http://dx.doi.org/10.1007/JHEP03(2022)043}{\emph{JHEP} {\bf 03} (2022)
  043}, [\href{https://arxiv.org/abs/2108.01105}{{\tt 2108.01105}}].

\bibitem{Suh:2021hef}
M.~Suh, \emph{{M2-branes wrapped on a topological disk}},
  \href{http://dx.doi.org/10.1007/JHEP09(2022)048}{\emph{JHEP} {\bf 09} (2022)
  048}, [\href{https://arxiv.org/abs/2109.13278}{{\tt 2109.13278}}].

\bibitem{Suh:2021aik}
M.~Suh, \emph{{D4-branes wrapped on a topological disk}},
  \href{http://dx.doi.org/10.1007/JHEP06(2023)008}{\emph{JHEP} {\bf 06} (2023)
  008}, [\href{https://arxiv.org/abs/2108.08326}{{\tt 2108.08326}}].

\bibitem{Karndumri:2022wpu}
P.~Karndumri and P.~Nuchino, \emph{{Five-branes wrapped on topological disks
  from 7D N=2 gauged supergravity}},
  \href{http://dx.doi.org/10.1103/PhysRevD.105.066010}{\emph{Phys. Rev. D} {\bf
  105} (2022) 066010}, [\href{https://arxiv.org/abs/2201.05037}{{\tt
  2201.05037}}].

\bibitem{Gutperle:2022pgw}
M.~Gutperle and N.~Klein, \emph{{A note on co-dimension 2 defects in N=4,d=7
  gauged supergravity}},
  \href{http://dx.doi.org/10.1016/j.nuclphysb.2022.115969}{\emph{Nucl. Phys. B}
  {\bf 984} (2022) 115969}, [\href{https://arxiv.org/abs/2203.13839}{{\tt
  2203.13839}}].

\bibitem{Suh:2022olh}
M.~Suh, \emph{{M5-branes and D4-branes wrapped on a direct product of spindle
  and Riemann surface}},  \href{https://arxiv.org/abs/2207.00034}{{\tt
  2207.00034}}.

\bibitem{Cheung:2022ilc}
K.~C.~M. Cheung, J.~H.~T. Fry, J.~P. Gauntlett and J.~Sparks, \emph{{M5-branes
  wrapped on four-dimensional orbifolds}},
  \href{http://dx.doi.org/10.1007/JHEP08(2022)082}{\emph{JHEP} {\bf 08} (2022)
  082}, [\href{https://arxiv.org/abs/2204.02990}{{\tt 2204.02990}}].

\bibitem{Couzens:2022lvg}
C.~Couzens, H.~Kim, N.~Kim, Y.~Lee and M.~Suh, \emph{{D4-branes wrapped on
  four-dimensional orbifolds through consistent truncation}},
  \href{http://dx.doi.org/10.1007/JHEP02(2023)025}{\emph{JHEP} {\bf 02} (2023)
  025}, [\href{https://arxiv.org/abs/2210.15695}{{\tt 2210.15695}}].

\bibitem{Faedo:2022rqx}
F.~Faedo, A.~Fontanarossa and D.~Martelli, \emph{{Branes wrapped on orbifolds
  and their gravitational blocks}},
  \href{http://dx.doi.org/10.1007/s11005-023-01671-1}{\emph{Lett. Math. Phys.}
  {\bf 113} (2023) 51}, [\href{https://arxiv.org/abs/2210.16128}{{\tt
  2210.16128}}].

\bibitem{Boido:2022iye}
A.~Boido, J.~P. Gauntlett, D.~Martelli and J.~Sparks, \emph{{Entropy Functions
  For Accelerating Black Holes}},
  \href{http://dx.doi.org/10.1103/PhysRevLett.130.091603}{\emph{Phys. Rev.
  Lett.} {\bf 130} (2023) 091603},
  [\href{https://arxiv.org/abs/2210.16069}{{\tt 2210.16069}}].

\bibitem{Boido:2022mbe}
A.~Boido, J.~P. Gauntlett, D.~Martelli and J.~Sparks, \emph{{Gravitational
  Blocks, Spindles and GK Geometry}},
  \href{https://arxiv.org/abs/2211.02662}{{\tt 2211.02662}}.

\bibitem{Hosseini:2019iad}
S.~M. Hosseini, K.~Hristov and A.~Zaffaroni, \emph{{Gluing gravitational blocks
  for AdS black holes}},
  \href{http://dx.doi.org/10.1007/JHEP12(2019)168}{\emph{JHEP} {\bf 12} (2019)
  168}, [\href{https://arxiv.org/abs/1909.10550}{{\tt 1909.10550}}].

\bibitem{Maldacena:1997re}
J.~M. Maldacena, \emph{{The Large N limit of superconformal field theories and
  supergravity}}, \href{http://dx.doi.org/10.1023/A:1026654312961}{\emph{Adv.
  Theor. Math. Phys.} {\bf 2} (1998) 231--252},
  [\href{https://arxiv.org/abs/hep-th/9711200}{{\tt hep-th/9711200}}].

\bibitem{Khavaev:1998fb}
A.~Khavaev, K.~Pilch and N.~P. Warner, \emph{{New vacua of gauged N=8
  supergravity in five-dimensions}},
  \href{http://dx.doi.org/10.1016/S0370-2693(00)00795-4}{\emph{Phys. Lett. B}
  {\bf 487} (2000) 14--21}, [\href{https://arxiv.org/abs/hep-th/9812035}{{\tt
  hep-th/9812035}}].

\bibitem{Freedman:1999gp}
D.~Z. Freedman, S.~S. Gubser, K.~Pilch and N.~P. Warner, \emph{{Renormalization
  group flows from holography supersymmetry and a c theorem}},
  \href{http://dx.doi.org/10.4310/ATMP.1999.v3.n2.a7}{\emph{Adv. Theor. Math.
  Phys.} {\bf 3} (1999) 363--417},
  [\href{https://arxiv.org/abs/hep-th/9904017}{{\tt hep-th/9904017}}].

\bibitem{Pilch:2000fu}
K.~Pilch and N.~P. Warner, \emph{{N=1 supersymmetric renormalization group
  flows from IIB supergravity}},
  \href{http://dx.doi.org/10.4310/ATMP.2000.v4.n3.a5}{\emph{Adv. Theor. Math.
  Phys.} {\bf 4} (2002) 627--677},
  [\href{https://arxiv.org/abs/hep-th/0006066}{{\tt hep-th/0006066}}].

\bibitem{Leigh:1995ep}
R.~G. Leigh and M.~J. Strassler, \emph{{Exactly marginal operators and duality
  in four-dimensional N=1 supersymmetric gauge theory}},
  \href{http://dx.doi.org/10.1016/0550-3213(95)00261-P}{\emph{Nucl. Phys. B}
  {\bf 447} (1995) 95--136}, [\href{https://arxiv.org/abs/hep-th/9503121}{{\tt
  hep-th/9503121}}].

\bibitem{Warner:1983vz}
N.~P. Warner, \emph{{Some New Extrema of the Scalar Potential of Gauged $N=8$
  Supergravity}},
  \href{http://dx.doi.org/10.1016/0370-2693(83)90383-0}{\emph{Phys. Lett. B}
  {\bf 128} (1983) 169--173}.

\bibitem{Warner:1983du}
N.~P. Warner, \emph{{Some Properties of the Scalar Potential in Gauged
  Supergravity Theories}},
  \href{http://dx.doi.org/10.1016/0550-3213(84)90286-4}{\emph{Nucl. Phys. B}
  {\bf 231} (1984) 250--268}.

\bibitem{Benna:2008zy}
M.~Benna, I.~Klebanov, T.~Klose and M.~Smedback, \emph{{Superconformal
  Chern-Simons Theories and AdS(4)/CFT(3) Correspondence}},
  \href{http://dx.doi.org/10.1088/1126-6708/2008/09/072}{\emph{JHEP} {\bf 09}
  (2008) 072}, [\href{https://arxiv.org/abs/0806.1519}{{\tt 0806.1519}}].

\bibitem{Klebanov:2008vq}
I.~Klebanov, T.~Klose and A.~Murugan, \emph{{AdS(4)/CFT(3) Squashed, Stretched
  and Warped}},
  \href{http://dx.doi.org/10.1088/1126-6708/2009/03/140}{\emph{JHEP} {\bf 03}
  (2009) 140}, [\href{https://arxiv.org/abs/0809.3773}{{\tt 0809.3773}}].

\bibitem{Ahn:2000aq}
C.-h. Ahn and J.~Paeng, \emph{{Three-dimensional SCFTs, supersymmetric domain
  wall and renormalization group flow}},
  \href{http://dx.doi.org/10.1016/S0550-3213(00)00687-8}{\emph{Nucl. Phys. B}
  {\bf 595} (2001) 119--137}, [\href{https://arxiv.org/abs/hep-th/0008065}{{\tt
  hep-th/0008065}}].

\bibitem{Ahn:2000mf}
C.-h. Ahn and K.~Woo, \emph{{Supersymmetric domain wall and RG flow from
  4-dimensional gauged N=8 supergravity}},
  \href{http://dx.doi.org/10.1016/S0550-3213(01)00008-6}{\emph{Nucl. Phys. B}
  {\bf 599} (2001) 83--118}, [\href{https://arxiv.org/abs/hep-th/0011121}{{\tt
  hep-th/0011121}}].

\bibitem{Corrado:2001nv}
R.~Corrado, K.~Pilch and N.~P. Warner, \emph{{An N=2 supersymmetric membrane
  flow}}, \href{http://dx.doi.org/10.1016/S0550-3213(02)00134-7}{\emph{Nucl.
  Phys. B} {\bf 629} (2002) 74--96},
  [\href{https://arxiv.org/abs/hep-th/0107220}{{\tt hep-th/0107220}}].

\bibitem{Aharony:2008ug}
O.~Aharony, O.~Bergman, D.~L. Jafferis and J.~Maldacena, \emph{{N=6
  superconformal Chern-Simons-matter theories, M2-branes and their gravity
  duals}}, \href{http://dx.doi.org/10.1088/1126-6708/2008/10/091}{\emph{JHEP}
  {\bf 10} (2008) 091}, [\href{https://arxiv.org/abs/0806.1218}{{\tt
  0806.1218}}].

\bibitem{Bobev:2009ms}
N.~Bobev, N.~Halmagyi, K.~Pilch and N.~P. Warner, \emph{{Holographic, N=1
  Supersymmetric RG Flows on M2 Branes}},
  \href{http://dx.doi.org/10.1088/1126-6708/2009/09/043}{\emph{JHEP} {\bf 09}
  (2009) 043}, [\href{https://arxiv.org/abs/0901.2736}{{\tt 0901.2736}}].

\bibitem{Bobev:2018wbt}
N.~Bobev, V.~S. Min, K.~Pilch and F.~Rosso, \emph{{Mass Deformations of the
  ABJM Theory: The Holographic Free Energy}},
  \href{http://dx.doi.org/10.1007/JHEP03(2019)130}{\emph{JHEP} {\bf 03} (2019)
  130}, [\href{https://arxiv.org/abs/1812.01026}{{\tt 1812.01026}}].

\bibitem{Bobev:2018uxk}
N.~Bobev, V.~S. Min and K.~Pilch, \emph{{Mass-deformed ABJM and black holes in
  AdS$_{4}$}}, \href{http://dx.doi.org/10.1007/JHEP03(2018)050}{\emph{JHEP}
  {\bf 03} (2018) 050}, [\href{https://arxiv.org/abs/1801.03135}{{\tt
  1801.03135}}].

\bibitem{deWit:1982bul}
B.~de~Wit and H.~Nicolai, \emph{{N=8 Supergravity}},
  \href{http://dx.doi.org/10.1016/0550-3213(82)90120-1}{\emph{Nucl. Phys. B}
  {\bf 208} (1982) 323}.

\bibitem{Cremmer:1978km}
E.~Cremmer, B.~Julia and J.~Scherk, \emph{{Supergravity Theory in
  Eleven-Dimensions}},
  \href{http://dx.doi.org/10.1016/0370-2693(78)90894-8}{\emph{Phys. Lett. B}
  {\bf 76} (1978) 409--412}.

\bibitem{deWit:1986oxb}
B.~de~Wit and H.~Nicolai, \emph{{The Consistency of the S**7 Truncation in D=11
  Supergravity}},
  \href{http://dx.doi.org/10.1016/0550-3213(87)90253-7}{\emph{Nucl. Phys. B}
  {\bf 281} (1987) 211--240}.

\bibitem{Bobev:2010ib}
N.~Bobev, N.~Halmagyi, K.~Pilch and N.~P. Warner, \emph{{Supergravity
  Instabilities of Non-Supersymmetric Quantum Critical Points}},
  \href{http://dx.doi.org/10.1088/0264-9381/27/23/235013}{\emph{Class. Quant.
  Grav.} {\bf 27} (2010) 235013}, [\href{https://arxiv.org/abs/1006.2546}{{\tt
  1006.2546}}].

\bibitem{Jafferis:2011zi}
D.~L. Jafferis, I.~R. Klebanov, S.~S. Pufu and B.~R. Safdi, \emph{{Towards the
  F-Theorem: N=2 Field Theories on the Three-Sphere}},
  \href{http://dx.doi.org/10.1007/JHEP06(2011)102}{\emph{JHEP} {\bf 06} (2011)
  102}, [\href{https://arxiv.org/abs/1103.1181}{{\tt 1103.1181}}].

\bibitem{Benini:2015noa}
F.~Benini and A.~Zaffaroni, \emph{{A topologically twisted index for
  three-dimensional supersymmetric theories}},
  \href{http://dx.doi.org/10.1007/JHEP07(2015)127}{\emph{JHEP} {\bf 07} (2015)
  127}, [\href{https://arxiv.org/abs/1504.03698}{{\tt 1504.03698}}].

\bibitem{Benini:2015eyy}
F.~Benini, K.~Hristov and A.~Zaffaroni, \emph{{Black hole microstates in
  AdS$_{4}$ from supersymmetric localization}},
  \href{http://dx.doi.org/10.1007/JHEP05(2016)054}{\emph{JHEP} {\bf 05} (2016)
  054}, [\href{https://arxiv.org/abs/1511.04085}{{\tt 1511.04085}}].

\bibitem{Behrndt:1996hu}
K.~Behrndt, R.~Kallosh, J.~Rahmfeld, M.~Shmakova and W.~K. Wong, \emph{{STU
  black holes and string triality}},
  \href{http://dx.doi.org/10.1103/PhysRevD.54.6293}{\emph{Phys. Rev. D} {\bf
  54} (1996) 6293--6301}, [\href{https://arxiv.org/abs/hep-th/9608059}{{\tt
  hep-th/9608059}}].

\bibitem{Duff:1999gh}
M.~J. Duff and J.~T. Liu, \emph{{Anti-de Sitter black holes in gauged N = 8
  supergravity}},
  \href{http://dx.doi.org/10.1016/S0550-3213(99)00299-0}{\emph{Nucl. Phys. B}
  {\bf 554} (1999) 237--253}, [\href{https://arxiv.org/abs/hep-th/9901149}{{\tt
  hep-th/9901149}}].

\bibitem{Cvetic:1999xp}
M.~Cvetic, M.~J. Duff, P.~Hoxha, J.~T. Liu, H.~Lu, J.~X. Lu et~al.,
  \emph{{Embedding AdS black holes in ten-dimensions and eleven-dimensions}},
  \href{http://dx.doi.org/10.1016/S0550-3213(99)00419-8}{\emph{Nucl. Phys. B}
  {\bf 558} (1999) 96--126}, [\href{https://arxiv.org/abs/hep-th/9903214}{{\tt
  hep-th/9903214}}].

\bibitem{Anabalon:2022aig}
A.~Anabal\'on, A.~Gallerati, S.~Ross and M.~Trigiante, \emph{{Supersymmetric
  solitons in gauged $ \mathcal{N} $ = 8 supergravity}},
  \href{http://dx.doi.org/10.1007/JHEP02(2023)055}{\emph{JHEP} {\bf 02} (2023)
  055}, [\href{https://arxiv.org/abs/2210.06319}{{\tt 2210.06319}}].

\end{thebibliography}\endgroup

\end{document}